\documentclass[aps,prb,twocolumn,nofootinbib,longbibliography,superscriptaddress]{revtex4-2}
\usepackage{amsfonts}
\usepackage{amscd}
\usepackage{amsmath}
\usepackage{amssymb}
\usepackage{newtxtext,newtxmath}
\usepackage{here}
\usepackage[dvipdfmx]{graphicx}
\usepackage{dcolumn}
\usepackage{bm}
\usepackage{xcolor}
\usepackage[%
  colorlinks=true,
  urlcolor=blue,
  linkcolor=blue,
  citecolor=blue
]{hyperref}
\usepackage{tabularx}
\usepackage{etoolbox}
\usepackage{braket}
\usepackage[caption=false]{subfig}
\newcommand{\nn}{\nonumber}
\newcommand{\beq}{\begin{equation}}
\newcommand{\eeq}{\end{equation}}

\makeatletter
\let\cat@comma@active\@empty
\makeatother

\allowdisplaybreaks

\usepackage[normalem]{ulem}

\renewcommand\sout{\bgroup \color{blue} \ULdepth=-.5ex \ULset}

\usepackage{hyperref}

\begin{document}

\title{
Odd-frequency pairs and anomalous proximity effect in nematic and chiral states of  \\
superconducting topological insulators
}

\author{Takeshi Mizushima}
\affiliation{Department of Materials Engineering Science, Osaka University, Toyonaka, Osaka 560-8531, Japan}
\author{Shun Tamura}
\affiliation{Department of Applied Physics, Nagoya University,
Nagoya 464-8603, Japan}
\author{Keiji Yada}
\affiliation{Department of Applied Physics, Nagoya University,
Nagoya 464-8603, Japan}
\author{Yukio Tanaka}
\affiliation{Department of Applied Physics, Nagoya University,
Nagoya 464-8603, Japan}

\date{\today}

\begin{abstract}

We investigate emergent odd-frequency pairs and proximity effect in nematic and chiral states of superconducting topological insulators (STIs), such as $M_x$Bi$_2$Se$_3$ ($M=$ Cu, Sr, Nb). The interplay of superconducting gap symmetry, the orbital degrees of freedom, and strong spin-orbit interaction generates a variety of odd-frequency pairs in the bulk and surface of STIs. The nematic and chiral states are the prototypes of topological superconductors with and without time-reversal symmetry, respectively. We find that the Fermi surface evolution from a closed spheroidal to an open cylindrical shape gives rise to the evolution of the emergent odd-frequency pairs and surface Andreev bound states (SABSs). In addition, spin polarization of odd-frequency pairs and SABSs stems from the non-unitary pairing in the chiral state. These evolution and spin polarization of odd-frequency pairs and SABSs can be captured by tunnel conductance spectroscopy. Furthermore, we study the anomalous proximity effect in various irreducible representations of STIs. The anomalous proximity effect is originally predicted in spin-triplet superconductor junctions without spin-orbit interaction; Odd-frequency spin-triplet $s$-wave pairs penetrate into diffusive normal metals (DN) and induce a pronounced zero-energy peak of the local density of states in the DN region. Here we demonstrate that contrary to the well-known results, the anomalous proximity effect in STIs is not immune to nonmagnetic impurities. The fragility is attributed to the fact that the proximitized odd-frequency even-parity pairs are admixtures of $s$-wave and non-$s$-wave pairs due to strong spin-orbit interaction inherent to the parent materials.
\end{abstract}

\maketitle

\section{Introduction}

The concept of odd-frequency pairs~\cite{ber74,tanaka12,escRPP15,linderRMP19} has paved the way for understanding anomalous equilibrium and nonequilibrium phenomena of superconductors (SCs). According to the Fermi-Dirac statistics, Cooper pair amplitude is an even or odd function in the relative time, leading to the notion of even-frequency pairing or odd-frequency pairing, respectively. Odd-frequency superconductivity, which is the superconducting state with odd-frequency gap function, has been discussed to occur in bulk electron systems, while its realization remains elusive~\cite{kirPRL91,balPRB92,eme92,col93,abrahamsPRB95,voj99,bel99,fus03,sol09,hot09,shi09,shi11,kus11,kus12,mat13,fom15,hos15,mat21}. 
%
Here we focus on the boundary of superconductors, which provides another platform to host odd-frequency pairs. The prototypical examples include the SC/ferromagnet junctions and SC/normal metal junctions~\cite{linderRMP19,berPRL01,berRMP05,esc03,esc08,esc11,escRPP15,tanPRL07,tanPRB07,esc07}. In the former case the broken spin-rotation symmetry convert even-frequency spin-singlet Cooper pairs into odd-frequency spin-triplet pairs~\cite{linderRMP19,berPRL01,berRMP05,esc03,esc08,esc11,escRPP15}. In the latter case, the translational symmetry breaking induces odd-frequency even-parity (odd-parity) pairs in the interface of an odd-parity (even-parity) SC and normal metal~\cite{tanPRL07,tanPRB07,esc07}. Recent progress on the understanding of odd-frequency pairing has shed light on the special roles of Andreev bound states (ABSs) in spin/charge transport, paramagnetic electromagnetic responses, and proximity effects~\cite{hig97,walter98,tanakaPRB05,tanPRB04,tanPRB05,tanPRL06,tanPRL07,tan09,asanoPRL11,yokPRL11,higashitaniPRB14,asanoPRB14,suzPRB14,suzPRB15,leePRB17,tanakaPRB18,tamPRB19,tan22,meissner,krieger,rogers,perrin,khaire,zdravkov,iovan,lenk,kapran,chi21}.

It has recently been pointed out that even in the absence of any symmetry breaking fields, odd-frequency pairing can emerge in the bulk of multiband SCs~\cite{blackshafferPRB13,komendovaPRB15,asanoPRB15,triola,tamura20,kanasugi} and double nanowires~\cite{ebisu16,triolaPRB19}. Triola {\it et al.} presented the general condition for the emergence of odd-frequency pairing~\cite{triola}, 
\beq
\sum _k \left[h_{nk}\Delta_{km}-\Delta_{nk}h^{\ast}_{km}\right] \neq 0,
\label{eq:OF}
\eeq
where ${h}_{nm}$ and ${\Delta}_{nm}$ are the normal state Hamiltonian and pair potential, respectively, and the indices represent spin, lattice, and band/orbital degrees of freedom. This condition enables to diagnose the symmetry and origin of  odd-frequency Cooper pairs. Equation~\eqref{eq:OF} is also associated with the concept called {\it superconducting fitness}~\cite{ramires16,ramires19} that reflects the compatibility of superconducting gap symmetry with a normal state Hamiltonian and measures the suppression of the superconducting critical temperature by symmetry breaking fields.

Another remarkable ingredient in SCs is a Majorana fermion, which is an elusive quasiparticle residing in vortices, surfaces, and interfaces~\cite{Schnyder08,qi09}. Topological SCs with chiral symmetry host Majorana fermions which exhibit uniaxial response~\cite{chuPRL09,nagJPSJ09,satoPRB2009,shiPRB10,mizushimaPRL12,mizushimaNJP13,tsutsumiJPSJ13,shiozakiPRB14,xiong}, multipole response~\cite{kobayashiPRL19,yamazaki}, and non-abelian statistics~\cite{uenoPRL13,satoPE14,liuPRX14,gaoPRB16,tanaka22}. 
Recently, the concept of odd-frequency pairs has encountered topological SCs. An important observation is the relationship between  zero energy states and odd-frequency  pairing~\cite{dainoPRB12,higashitaniPRB12,asanoPRB13,mizushimaPRB14v2,tamuraPRB19,stanev,tsi19,cay20,kuz20}. In chiral symmetric systems, the topological invariant not only counts the number of the zero energy states but also determines the spectral property of the odd-frequency pairs at the zero frequency limit~\cite{tamuraPRB19}. This is called the {\it spectral bulk-boundary correspondence} (SBBC), which is a generalization of bulk-boundary correspondence into complex frequencies~\cite{tamuraPRB19,daiPRB19,tanaka21,tamura21}. The SBBC was recently proved by Daido and Yanase~\cite{daiPRB19}, by introducing the notion of chirality polarization.
In analogy with the bulk-boundary correspondence of electric polarization that is the relation between the amount of surface-accumulated electric charge and a geometric Berry phase defined in the bulk, the SBBC refers to the equivalence of chirality charge with chirality polarization, where the chirality charge represents the surface-accumulated odd-frequency even-parity pairs.

The superconducting topological insulators (STIs), such as $M_x$Bi$_2$Se$_3$ ($M=$ Cu, Sr, Nb), provide a platform for investigating Majorana fermions~\cite{satoPRB10,fuPRL10,sasakiPRL11,yamakage12,hao11,hsieh12,yipPRB13,mizushimaPRB14,sasakiPC15} and odd-frequency pairs. 
Several bulk measurements have revealed the emergence of the uniaxial anisotropy at superconducting critical temperature~\cite{matano,yonezawa,panSR16,nikitinPRB16,asabaPRX17,shenNPJ17,du17,smylie17,smylie18,kunNJP18,willa}. The rotation symmetry breaking is compatible with odd-parity time-reversal invariant pairing belonging to the $E_u$ irreducible representation of the point group $D_{3d}$, which exhibits twofold symmetric gap anisotropy~\cite{fuPRB14}. 
Concurrently with bulk measurements, surface sensitive probes were used to reveal the surface states. In point contact measurements for $M={\rm Cu}$, the zero bias conductance peak (ZBCP) was reported as a first evidence of topological superconductivity~\cite{sasakiPRL11}. Although the similar ZBCPs have been observed independently~\cite{kir12,pen13}, there has been the conflicting report~\cite{pen13}. The significant enhancement of zero bias conductance has been observed in $M={\rm Nb}$~\cite{kur19}.  Scanning tunneling microscopy/spectroscopy (STM/STS) showed a fully gapped local density of states at the Fermi level reminiscent of a conventional superconducting gap~\cite{levyPRL13}, though the vortices are uniaxially elongated and host ZBCPs consistent with nematic and topological superconductivity~\cite{tao18}. No ZBCP has also been observed by STM/STS in $M={\rm Sr}$~\cite{han15,du17,kum21,bag22} and $M={\rm Nb}$~\cite{sir18,wil18}. Hence, the surface states of $M_x$Bi$_2$Se$_3$ still remain controversial.

Another important feature is that chemical dopants intercalate between the quintuple layers of the parent material, involving the Fermi surface evolution from a closed ellipsoidal shape to an open cylindrical shape. This may drive multiple phase transitions between competing superconducting states~\cite{uematsu,yuanPRB17,akz20}. When the Fermi surface is cylindrical, the chiral state with spontaneously broken time-reversal symmetry, which belongs to the $E_u$ representation of $D_{3d}$, is fully gapped and gains the condensation energy. Indeed, the in-plane $H_{\rm c2}$ anisotropy in Cu$_x$Bi$_2$Se$_3$ disappears with increasing the carrier density ($x=0.46$ and $0.54$), indicating that the superconducting gap becomes isotropic~\cite{kawai}. Although no evidence of the time-reversal symmetry breaking has been reported in $M_x$Bi$_2$Se$_3$ so far, it is important to investigate the theory of anomalous proximity effect in order to discriminate a signal of the chiral state from other pairing states.

In this paper, we investigate emergent odd-frequency pairs and anomalous proximity effect in the nematic and chiral states of STIs.  
These materials possess several peculiarities; (i) the orbital degrees of freedom, (ii) strong spin-orbit coupling, (iii) Majorana fermions, and (iv) topological phase transition. 
We first demonstrate that the first two factors (i,ii) enrich the properties of odd-frequency Cooper pairs in the bulk of $M_x$Bi$_2$Se$_3$.
Then, we categorize emergent pair amplitudes on the surface and junctions of STIs in terms of irreducible representations of $D_{3d}$ symmetry. We clarify that nematic and chiral states are accompanied by helical and spin-polarized Majorana fermions, respectively, which can be captured by the characteristic spectra of tunneling conductance. 
As mentioned above, the recent STM/STS measurements have not observed pronounced ZBCP. As an alternative way to discriminate the nematic and chiral states from the conventional pairing state, we 
%
examine the anomalous proximity effect in the junction of STI and dirty normal metals (DN). In the absence of spin-orbit interaction, odd-frequency $s$-wave pairs emerge in the interface of spin-triplet odd-parity SC and DN and penetrate into DN. As $s$-wave pairs are tolerant to nonmagnetic impurities, the odd-frequency $s$-wave pairs proximitized to the DN side is responsible for a pronounced ZBCP as a signature of odd-parity superconductivity~\cite{tanPRB04,tanPRB05,tanPRL06,ikePRB16}. Contrary to such well-known results, we find that the anomalous proximity effect in STI/DN junctions is sensitive to the strength of nonmagnetic impurities in the DN region even through odd-frequency even-parity pairs appear in the interface region. The fragility is attributed to the interplay of gap symmetry, the orbital degrees of freedom, and strong spin-orbit interaction.

The organization of this paper is as follows. In Sec.~II, we start with the symmetry classification of even- and odd-frequency pair amplitudes. We describe the Hamiltonian of the parent material and discuss the emergent Cooper pairs in bulk STIs. In Sec.~III, we present the numerical results of surface ABSs and odd-frequency pairs in the nematic and chiral states. In Sec.~IV, we present the tunneling conductance in the junction of STIs, which captures the characteristic dispersion of low-lying ABSs. We also discuss the anomalous proximity effect in STI/DN junctions. Section~V is devoted to a summary and conclusion. 
In Appendix A, we describe the calculation of the critical temperatures in each irreducible representation. We show the change of the carrier density and the hexagonal warping drives the nematic-to-chiral phase transition. In Appendix B, we demonstrate that nematic states obey the SBBC and topological criticality. The anomalous proximity effect in Dirac SCs without spin-orbit coupling is discussed in Appendic C. In this paper, we introduce the Pauli matrices in the spin, orbital, and particle-hole spaces, $s_{\mu}$, $\hat{\sigma}_{\mu}$, and $\check{\tau}_{\mu}$, respectively, where $s_0$, $\hat{\sigma}_0$, and $\check{\tau}_0$ are the unit matrices in each space. We also set $\hbar=k_{\rm B}=1$.

\section{Classification of Cooper pair amplitudes}

\subsection{Cooper pair amplitudes}
\label{sec:bulkOF}

Here we classify the Cooper pair amplitudes emergent in SCs with two orbital degrees of freedom, where the orbital and spin degrees of freedom are denoted by $\sigma = 1, 2$ and $s=\uparrow,\downarrow$, respectively. Let us start with 
the Matsubara Green's function,
\begin{align}
\check{G}({\bm r}_1,{\bm r}_2,i\varepsilon_n) &= - \int^{1/T}_0 d\tau e^{i\varepsilon_n\tau} \left\langle
T_{\tau}{\bm \Psi}({\bm r}_1,\tau) \bar{\bm \Psi}({\bm r}_2,0) 
\right\rangle \nn \\
&\equiv \begin{pmatrix}
\hat{G}({\bm r}_1,{\bm r}_2,i\varepsilon _n) & \hat{F}({\bm r}_1,{\bm r}_2,i\varepsilon _n) \\
\hat{\bar{F}}({\bm r}_1,{\bm r}_2,i\varepsilon _n) & \hat{\bar{G}}({\bm r}_1,{\bm r}_2,i\varepsilon _n),
\end{pmatrix}
\label{eq:Gorkov}
\end{align}
where $\varepsilon_n=(2n+1)\pi T$ is the Matsubara frequency at temperature $T$ ($n\in\mathbb{Z}$). The field operator in the Nambu space is defined as ${\bm \Psi}^{\rm tr} = [c_{(\uparrow,1)},c_{(\downarrow,1)},c_{(\uparrow,2)},c_{(\downarrow,2)},c^{\dag}_{(\uparrow,1)},c^{\dag}_{(\downarrow,1)},c^{\dag}_{(\uparrow,2)},c^{\dag}_{(\downarrow,2)}]^{\rm tr}$, where $c_{\alpha}\equiv c_{\alpha}({\bm r},\tau)$ is the annihilation operator of an electron with $\alpha  \equiv (s,\sigma)$ and $a^{\rm tr}$ is the transpose of a matrix $a$. The Green's function for clean systems is obtained by solving the Gor'kov equation
\beq
{\check{G}}({\bm r}_1,{\bm r}_2,i\varepsilon _n) = \left[
i\varepsilon _n - \check{\mathcal{H}}({\bm r}_1,{\bm r}_2)
\right]^{-1} ,
\label{eq:G}
\eeq
where $\check{\mathcal{H}}({\bm r}_1,{\bm r}_2)$ is the Bogoliubov-de Gennes (BdG) Hamiltonian density in the particle-hole space and its explicit form is given in Eq.~\eqref{eq:Hbdg}.

The anomalous component of the Green's function in the Nambu space, $\hat{F}$, which represents the Cooper pair amplitudes under the gap function $\hat{\Delta}$. Following the conventional terminology~\cite{tanaka12}, we classify the Cooper pair amplitudes in terms of the parities under the exchange of the imaginary time (Matsubara frequency), spin, the relative coordinate of the two electrons, and electron orbital, $(\eta_{\rm time}, \eta_{\rm spin}, \eta_{\rm mom}, \eta_{\rm orb})=(\pm 1,\pm 1,\pm 1, \pm 1)$, as 
\begin{align}
&\eta_{\rm time}F_{(s_1\sigma_1),(s_2\sigma_2)}({\bm r}_1,{\bm r}_2,i\varepsilon _n)
= F_{(s_1\sigma_1),(s_2\sigma_2)}({\bm r}_1,{\bm r}_2,-i\varepsilon _n), \label{eq:f1} \\
&\eta_{\rm spin}F_{(s_1\sigma_1),(s_2\sigma_2)}({\bm r}_1,{\bm r}_2,i\varepsilon _n)
= F_{(s_2\sigma_1),(s_1\sigma_2)}({\bm r}_1,{\bm r}_2,i\varepsilon _n), \label{eq:f2} \\
&\eta_{\rm mom}F_{(s_1\sigma_1),(s_2\sigma_2)}({\bm r}_1,{\bm r}_2,i\varepsilon _n)
= F_{(s_1\sigma_1),(s_2\sigma_2)}({\bm r}_2,{\bm r}_1,i\varepsilon _n), \label{eq:f3} \\
&\eta_{\rm orb}F_{(s_1\sigma_1),(s_2\sigma_2)}({\bm r}_1,{\bm r}_2,i\varepsilon _n)
= F_{(s_1\sigma_2),(s_2\sigma_1)}({\bm r}_1,{\bm r}_2,i\varepsilon _n). \label{eq:f4}
\end{align}
The Fermi-Dirac statistics requires the combination of the parities to satisfy the relation
\beq
\eta_{\rm time}\eta_{\rm spin}\eta_{\rm mom}\eta_{\rm orb} =-1,
\label{eq:Fsymmetry}
\eeq
or equivalently
$\hat{F}({\bm r}_1,{\bm r}_2,i\varepsilon_n) = - \hat{F}^{\rm tr}({\bm r}_2,{\bm r}_1,-i\varepsilon_n) $.

\begin{table}[t!]
\caption{Pairing symmetries in terms of parities under the exchange of the time, spin, spatial coordinate, and orbital indices of pairs, ($\eta_{\rm time}$, $\eta_{\rm spin}$, $\eta_{\rm mom}$, $\eta_{\rm orb}$) [See also Eqs.~\eqref{eq:f1}-\eqref{eq:f4}]. 
}
\begin{ruledtabular}
\begin{tabular}{ccccccc}
& category & $\eta_{\rm time}$ & $\eta_{\rm spin}$ & $\eta_{\rm mom}$ & $\eta_{\rm orb}$ & \\
\hline
&ESEE & $+$ & $-$ & $+$ & $+$ & \\
&ESOO & $+$ & $-$ & $-$ & $-$ & \\
&ETOE & $+$ & $+$ & $-$ & $+$ & \\
&ETEO & $+$ & $+$ & $+$ & $-$ & \\
\hline
&OSOE & $-$ & $-$ & $-$ & $+$ & \\
&OSEO & $-$ & $-$ & $+$ & $-$ & \\
&OTEE & $-$ & $+$ & $+$ & $+$ & \\
&OTOO & $-$ & $+$ & $-$ & $-$ & \\
\end{tabular}
\end{ruledtabular}
\label{table0}
\end{table}

In Table~\ref{table0}, we summarize the classes of the Cooper pair amplitudes. The Cooper pair amplitudes can be classified into eight-fold way in terms of ($\eta_{\rm time}$, $\eta_{\rm spin}$, $\eta_{\rm mom}$, $\eta_{\rm orb}$). The even- and odd-frequency pair amplitudes, $\hat{F}^{\rm even}(i\varepsilon_n)$ and $\hat{F}^{\rm odd}(i\varepsilon_n)$, are defined as
\begin{gather}
\hat{{F}}^{\rm even}(i\varepsilon_n) = \frac{1}{2}\left[
\hat{{F}}(i\varepsilon_n) + \hat{{F}}(-i\varepsilon_n)
\right], \\
\hat{{F}}^{\rm odd}(i\varepsilon_n) = \frac{1}{2}\left[
\hat{{F}}(i\varepsilon_n) - \hat{{F}}(-i\varepsilon_n)
\right],
\end{gather}
where $\hat{{F}}^{\rm even}$ and $\hat{{F}}^{\rm odd}$ belong to the class of $\eta_{\rm time}=+1$ and $\eta_{\rm time}=-1$, respectively. The classes with $\eta_{\rm time}=+1$ include the even-frequency spin-singlet even-parity even-orbital (ESEE) class, the even-frequency spin-singlet odd-parity odd-orbital (ESOO) class, the even-frequency spin-triplet odd-parity even-orbital (ETOE) class, and the even-frequency spin-triplet even-parity odd-orbital (ETEO) class. The odd-frequency pairs with $\eta _{\rm time}=-1$ are categorized to the odd-frequency spin-singlet odd-parity even-orbital (OSOE) class, the odd-frequency spin-singlet even-parity odd-orbital (OSEO) class, the odd-frequency spin-triplet even-parity even-orbital (OTEE) class, and the odd-frequency spin-triplet odd-parity odd-orbital (OTOO) class. 


The Green's function in the clean limit is determined by the BdG Hamiltonian,
\beq
\check{\mathcal{H}} = \begin{pmatrix}
\hat{h} & \hat{\Delta} \\ \hat{\Delta}^{\dag} & -\hat{h}^{\rm tr}
\end{pmatrix},
\label{eq:Htotal}
\eeq
where $\hat{h}$ and $\hat{\Delta}$ are the $N\times N$ matrix of the single-particle Hamiltonian and pair potential, respectively, and $N$ includes the spin and orbital degrees of freedom and the number of lattice sites. These submatrices obey the hermiticity $\hat{h}=\hat{h}^{\dag}$ and the Fermi statistics $\hat{\Delta}=-\hat{\Delta}^{\rm tr}$, leading to the particle-hole symmetry, $\check{\tau}_x\check{\mathcal{H}}^{\rm tr}\check{\tau}_x = - \check{\mathcal{H}}$. 

To elucidate the Cooper pair amplitudes, let us consider the weak coupling limit $\Delta \rightarrow 0$. The anomalous Green's function in the Matsubara representation, $\hat{F}(i\varepsilon_n)$, is obtained from the Gor'kov equation \eqref{eq:G} as
\begin{align}
\hat{F}(i\varepsilon_n)= -\left[1+\hat{G}_{\rm N}(i\varepsilon_n)\hat{\Delta}\hat{\bar{G}}_{\rm N}(i\varepsilon_n)\hat{\Delta}^{\dag}\right]^{-1}
\hat{G}_{\rm N}(i\varepsilon_n)\hat{\Delta}\hat{\bar{G}}_{\rm N}(i\varepsilon_n),
\end{align}
where $\hat{G}_{\rm N}(i\varepsilon_n)=(i\varepsilon_n - \hat{h})^{-1}$ and $\hat{\bar{G}}_{\rm N}(i\varepsilon_n)=(i\varepsilon_n + \hat{h}^{\rm tr})^{-1}$ are the Matsubara Green's function in the normal state. The leading order contribution of the odd-frequency pair amplitude is then given by~\cite{triola}
\beq
\hat{F}^{\rm odd}(i\varepsilon _n) 
= -i\varepsilon_n\left(\varepsilon^2_n+\hat{h}^2\right)^{-1}
\left[\hat{h} \hat{\Delta} -\hat{\Delta}\hat{h}^{\ast} \right]
\left(\varepsilon^2_n+\hat{h}^{\ast 2}\right)^{-1},
\label{eq:Fodd}
\eeq
and the even frequency pair amplitude is 
\beq
\hat{F}^{\rm even}(i\varepsilon _n) 
= \left(\varepsilon^2_n+\hat{h}^2\right)^{-1}
\hat{h} \hat{{\Delta}}\hat{h}^{\ast} 
\left(\varepsilon^2_n+\hat{h}^{\ast 2}\right)^{-1}.
\label{eq:evenF}
\eeq
Equation \eqref{eq:Fodd} indicates that the odd-frequency pairing can emerge unless 
\beq
\hat{h} \hat{\Delta} -\hat{\Delta}\hat{h}^{\ast}=0.
\label{eq:odd_condition}
\eeq
When the normal state maintains the time-reversal symmetry, i.e., $(is_y)\hat{h}^{\rm tr}(-is_y)=\hat{h}$, the inverse of the normal Green's function is given by $\hat{G}^{-1}_{\rm N}(i\varepsilon_n) = i\varepsilon_n - \hat{h}$ and $\hat{\bar{G}}_{\rm N}(i\varepsilon_n)=s_y\hat{G}_{\rm N}(-i\varepsilon_n)s_y$. 
Equation~\eqref{eq:Fodd} is then rewritten to
\beq
\hat{\mathcal{F}}^{\rm odd}(i\varepsilon _n) 
= -i\varepsilon_n\left(\varepsilon^2_n+\hat{h}^2\right)^{-1}
\left[\hat{h}, \hat{\tilde{\Delta}} \right]
\left(\varepsilon^2_n+\hat{h}^2\right)^{-1},
\label{eq:Fodd2}
\eeq
where we have introduced 
$\hat{\mathcal{F}} \equiv -i \hat{F}s_y$ and
$\hat{\tilde{\Delta}} \equiv -i \hat{\Delta} s_y$.

\subsection{Superconducting topological insulators}
\label{sec:model}

\begin{figure}[t]
\includegraphics[width=85mm]{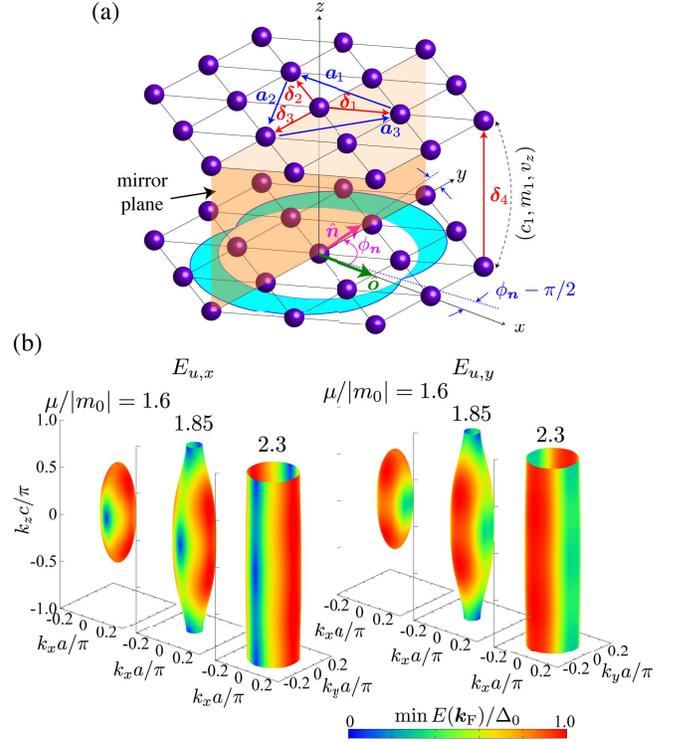}
\caption{(a) Crystal structure in Eq.~\eqref{eq:hti} and the gap structure of the $E_u$ nematic state. The crystal is composed of the hexagonal layers stacked along the $z$ axis and the nodal direction of the superconducting gap is related to the nematic angle $\phi_{\bm n}$. The shaded are in the $yz$ plane is one of the mirror planes maintained by Eq.~\eqref{eq:hti}. (b) Fermi surface and superconducting gap of the $E_{u,x}$ and $E_{u,y}$ nematic state (color map) for several values of the chemical potential, $\mu/|m_0|=1.6$, 1.85, and 2.3. In (b), we set $\lambda k^2_{{\rm F},\parallel}/v=0.1$.}
\label{fig:fs}
\end{figure}

In this paper, we consider the superconducting state of doped topological insulators, $M_x$Bi$_2$Se$_3$ ($M=$ Cu, Sr, Nb). The low-lying electrons have orbital degrees of freedom in addition to spin $1/2$, and the low-energy effective Hamiltonian is describable with two orbitals constituted from two $p_z$ orbitals localized on the lower and upper sides of the quintuple layer. The Hamiltonian for the parent topological insulators is given by~\cite{zhang2009,liuPRB10} \begin{align}
\hat{h}({\bm k}) =& c({\bm k}) -\mu + m({\bm k})\hat{\sigma}_x + v_z f_z({\bm k}) \hat{\sigma}_y 
+ v\left( {\bm f}({\bm k})\times {\bm s} \right)_z\hat{\sigma}_z \nn \\
&+ \lambda f_{3\lambda}({\bm k})s_z\hat{\sigma}_z .
\label{eq:hti}
\end{align}
where $\mu$ is the chemical potential. We have introduced $c ({\bm k}) = c_0 + c_1 f_{\perp}({\bm k})+c_2f_{\parallel}({\bm k})$ and $m ({\bm k}) = m_0 + m_1 f_{\perp}({\bm k}) + m_2f_{\parallel}({\bm k})$. We consider a tight-binding model on the hexagonal layers stacked along the $z$-axis~\cite{hashimotoJPSJ13,hashimoto14,haoPRB17},
\begin{align}
&f_x = [ \sin({\bm k}\cdot{\bm \delta}_1) - \sin({\bm k}\cdot{\bm \delta}_2) ]/(\sqrt{3}a), \\
&f_y = [ \sin({\bm k}\cdot{\bm \delta}_1) + \sin({\bm k}\cdot{\bm \delta}_2) - 2 \sin({\bm k}\cdot{\bm \delta}_3)]/(3a), \\
&f_z = \sin({\bm k}\cdot{\bm \delta}_4)/c, \\
&f_{\parallel}=\frac{4}{3a^2}\left[3-\cos({\bm k}\cdot{\bm \delta}_1)-\cos({\bm k}\cdot{\bm \delta}_2)-\cos({\bm k}\cdot{\bm \delta}_3)\right], \\
&f_{\perp} =2[1-\cos({\bm k}\cdot{\bm \delta}_4)]/c^2.
\end{align}
As shown in Fig.~\ref{fig:fs}(a), the nearest-neighbor bond vectors are defined as 
${\bm \delta}_1=(\frac{\sqrt{3}}{2}a,\frac{1}{2}a,0)$, ${\bm \delta}_2=(-\frac{\sqrt{3}}{2}a,\frac{1}{2}a,0)$, ${\bm \delta}_3=(0,-a,0)$, and ${\bm \delta}_4=(0,0,c)$, where $a$ and $c$ are in-plane and out-of-plane lattice constants, respectively. The third and fifth terms of the right-hand side in Eq.~\eqref{eq:hti} represent the insulating gap and the spin-orbit coupling, respectively. 
The last term in Eq.~\eqref{eq:hti} introduces three mirror planes and threefold rotational symmetry in the $xy$ plane and gives rise to the hexagonal warping of the axially symmetric Fermi surface~\cite{fuPRL09,fuPRB14},
\beq
f_{3\lambda}({\bm k}) = - \frac{16}{3\sqrt{3}a^3} 
\sum^{3}_{j=1}\sin {\bm k}\cdot{\bm a}_j .
\label{eq:warp}
\eeq
The next-nearest-neighbor bond vectors are introduced as ${\bm a}_1={\bm \delta}_1-{\bm \delta}_2$, ${\bm a}_2={\bm \delta}_2-{\bm \delta}_3$, and ${\bm a}_3={\bm \delta}_3-{\bm \delta}_1$.
In the vicinity of the $\Gamma$ point, the Hamiltonian in Eq.~\eqref{eq:hti} reduces to
$\hat{h}({\bm k}) \approx c({\bm k}) + m({\bm k})\sigma _x + v_z k_z \sigma _y 
+ v( {\bm k}\times {\bm s} )_z\sigma _z
+ \lambda (k^3_++k^3_-)s_z \sigma _z$,
with replacing $f_{\mu}({\bm k})$ to $k^2_{\mu}$ ($\mu = x, y,z$), where $k_{\pm}\!\equiv\!k_x\pm ik_y$. 
The quasiparticle states in STIs are obtained by diagonalizing the BdG Hamiltonian in Eq.~\eqref{eq:Htotal} with Eq.~\eqref{eq:hti}.

The $4\times 4$ Hamiltonian, $\hat{h}({\bm k})$, maintains the inversion symmetry and time-reversal symmetry
\begin{gather}
P\hat{h}({\bm k})P^{\dag} = \hat{h}(-{\bm k}),
\label{eq:inv} \\
\mathcal{T}\hat{h}({\bm k})\mathcal{T}^{-1} = \hat{h}(-{\bm k}), 
\label{eq:trs}
\end{gather}
where $P = \hat{\sigma}_x$ and $\mathcal{T} = is_y K$ ($K$ is the complex conjugation operator). In addition, $\hat{h}({\bm k})$ is invariant under a mirror reflection ($M$) and threefold rotation about the $\hat{\bm z}$-axis ($R_3$), 
\begin{gather}
{M}\hat{h}({\bm k}){M}^{\dag} = \hat{h}(\underline{\bm k}),
\label{eq:mirror} \\
U_3 \hat{h} ({\bm k}) U^{\dag}_3 = \hat{h}(R_3{\bm k}).
\label{eq:rot}
\end{gather}
The mirror operator, 
${M} = i{\bm s}\cdot\hat{\bm o}$,
flips the momentum ${\bm k}$ and the spin ${\bm s}$ to $\underline{\bm k}={\bm k}-2\hat{\bm o}({\bm k}\cdot\hat{\bm o})$ and $\underline{\bm s}=-{\bm s}+2\hat{\bm o}({\bm s}\cdot\hat{\bm o})$, where $\hat{\bm o}=(\cos\phi,\sin\phi,0)$ is three mirror axes of the crystal, $\phi  = 0, \pm 2\pi/3$ [see Fig.~\ref{fig:fs}(a)].


The parent material, Bi$_2$Se$_3$, has a layered structure constituted from stacked Se-Bi-Se-Bi-Se quintuple layers along the (111) direction ($z$-axis in this paper). The layers are weakly bounded by van der Waals forces. The intercalation of Cu atoms into the van der Waals gap between the quintuple layers of Bi$_2$Se$_3$ increases the carrier density in the conduction band and induces a small electrons pocket around the $\Gamma$ point. The highest $T_{\rm c}$ of Cu$_x$Bi$_2$Se$_3$, $T_{\rm c}\approx 3.8~{\rm K}$, is accomplished by the optimal doping $0.12<x<0.15$, where the electron carrier density is about $10^{19}$-$10^{20}~{\rm cm}^{-3}$. From the analysis based on the Shubnikov-de Haas measurement of Cu$_x$Bi$_2$Se$_3$, Lahoud {\it et al.}~\cite{lahoudPRB13} reported the Fermi surface evolution from a spheroidal to cylindrical shape around the $10^{20}~{\rm cm}^{-3}$ carrier density. Nontrivial bulk topological superconductivity is realized in odd parity pairs when the number of Fermi surfaces embracing time-reversal invariant momenta ($\Gamma$ and $Z$ points) is odd~\cite{fuPRL10,satoPRB10}. The Fermi surface evolution is therefore indispensable for fully understanding surface ABSs in this material. 

The Fermi surface evolution may be associated with the fact that the intercalation enlarges the van der Waals gap and alters the hopping energies along the $z$-direction, $(c_1,m_1,v_z)$, from those in the parent material. To incorporate this effect, we parameterize the hopping energies along the $c$-axis as $(c_1,m_1,v_z)=\alpha(\mu)(c^{(0)}_1,m^{(0)}_1,v^{(0)}_z)$. We use the values of $(c^{(0)}_1,m^{(0)}_1,v^{(0)}_z)$ and all other parameters of Eq.~\eqref{eq:hti} reported in Ref.~\cite{hashimoto14}. The strength of the hexagonal warping term in Eq.~\eqref{eq:hti} is scaled as $\lambda k^2_{{\rm F},\parallel}/v$, using the strength of the spin-orbit coupling $v$ and $k_{{\rm F},\parallel}\equiv \sqrt{(\mu^2-m^2_0)/v}$. The evolution of the Fermi surface topology is described by the single parameter $\alpha(\mu)$, where we take $\alpha=1.0$ at $\mu = 1.6|m_0|$ and $\alpha=0.3$ at $2.3|m_0|$, and the intermediate region is interpolated linearly. The $\mu$-dependence of the Fermi surface is displayed in Fig.~\ref{fig:fs}(b), where the Fermi surface topology changes at $\mu=1.8|m_0|$. The carrier density in the conduction band, $n _{\rm CB}$, changes from $n _{\rm CB}=8.4\times 10^{19}~{\rm cm}^{-3}$ at $\mu = 1.6|m_0|$ (spheroidal Fermi surface) to $n _{\rm CB}=5.5\times 10^{20}~{\rm cm}^{-3}$ at $\mu = 2.3|m_0|$ (cylindrical Fermi surface). 

\subsection{Nematic/chiral states and pair amplitudes in bulk STIs}
\label{sec:bulk}

Electrons in the parent material of STIs inevitably have the orbital degrees of freedom and strong spin-orbit coupling, which enables topological odd-parity pairs even in an $s$-wave channel.
We first consider even-frequency $s$-wave pairing in a $D_{3d}$ crystal. Let $\Gamma$ be the irreducible representations of $D_{3d}$ with the dimension $n_{\Gamma}$ and $\{\hat{d}^{\Gamma}_{1},\cdots,\hat{d}^{\Gamma}_{n_{\Gamma}}\}$ be basis functions of $\Gamma$. The pair amplitude is then expanded in terms of the basis functions as
\beq
\hat{F} ({\bm k},{\bm R},i\varepsilon_n) 
= i\sum^{n_{\Gamma}}_{j=1} \hat{\mathcal{F}} ({\bm R},i\varepsilon_n) \hat{d}^{\Gamma}_j({\bm k})s_y,
\label{eq:delta_orbital}
\eeq
where ${\bm R}$ is the center-of-mass coordinate of the pair amplitude.
In Table~\ref{table_D3d}, we summarize the possible basis functions. In centrosymmetric materials, the pair amplitudes obey the inversion symmetry, 
\beq
{P}\hat{F}({\bm k},{\bm R},i\varepsilon_n) {P}^{\dag}=\eta_{P} \hat{F}(-{\bm k},-{\bm R},i\varepsilon_n) , 
\label{eq:parity_delta}
\eeq
where $P\equiv \sigma_x$ and $\eta _{P} = \pm 1$.

\begin{table}[t!]
\caption{Pairing symmetries according to the irreducible representation of ${\tt D_{3d}}$ point group $\Gamma=\{A_{1g}^{({\rm e})}, A_{1u}^{({\rm e})}, A_{2u}^{({\rm e})}, E_u^{({\rm e})} \}$ for even frequency pairing and $\Gamma=\{A_{1g}^{({\rm o})}, A_{1u}^{({\rm o})}, A_{2u}^{({\rm o})}, E_u^{({\rm o})} \}$ for odd-frequency pairing, where $\mathcal{C}_3=e^{-i(\pi/3)s_z}$, $\mathcal{C}_2=e^{-i(\pi/2)s_y}\sigma _x$, and $\mathcal{M}=-is_x$ are the operators associated with three-fold rotation symmetry, two-fold rotation symmetry, and the mirror symmetry, respectively. }
\begin{ruledtabular}
\begin{tabular}{ccccccccc}
& category & basis ($\hat{d}^{\Gamma}_j$) & $\mathcal{C}_3$ & $\mathcal{C}_2$ & $\mathcal{M}$ & $\eta_{P}$ & $\Gamma$ & \\
\hline
&ESEE & \begin{tabular}{c} $\sigma _0,\sigma _x$ \\ $\sigma _z$ \end{tabular} & 
\begin{tabular}{c} $+1$ \\ $+1$\end{tabular} & \begin{tabular}{c} $+1$ \\ $-1$\end{tabular} & 
\begin{tabular}{c} $+1$ \\ $+1$\end{tabular} & \begin{tabular}{c} $+1$ \\ $-1$\end{tabular} & \begin{tabular}{c} $A^{({\rm e})}_{1g}$ \\ $A^{({\rm e})}_{2u}$ \end{tabular}& \\
\hline
&ETEO & \begin{tabular}{c} $s_z\sigma _y$ \\ $(s_x\sigma _y,s_y\sigma _y)$ \end{tabular} & 
\begin{tabular}{c} $+1$ \\ $-1$\end{tabular} & \begin{tabular}{c} $+1$ \\ $0$\end{tabular} & 
\begin{tabular}{c} $-1$ \\ $(+1,-1)$\end{tabular} & \begin{tabular}{c} $-1$ \\ $-1$\end{tabular} & \begin{tabular}{c} $A^{({\rm e})}_{1u}$ \\ $E^{({\rm e})}_u$ \end{tabular}& \\
\hline
&OSEO & $\sigma _y$ & $+1$ & $-1$ & $+1$ & $-1$ & $A^{({\rm o})}_{2u}$ &\\
\hline
&OTEE & \begin{tabular}{c} $s_z\sigma_0, s_z\sigma _x$ \\ $(s_x,s_y)\sigma_0$, $(s_x, s_y)\sigma_x$ \\ $s_z\sigma _z$ \\ $(s_x\sigma _z,s_y\sigma _z)$ \end{tabular} &
\begin{tabular}{c} $+1$ \\ $-1$ \\ $+1$ \\ $-1$ \end{tabular} & \begin{tabular}{c} $-1$ \\ $0$ \\ $+1$ \\ $0$ \end{tabular} & 
\begin{tabular}{c} $-1$ \\ $(+1,-1)$ \\ $-1$ \\ $(+1,-1)$ \end{tabular} & \begin{tabular}{c} $+1$ \\ $+1$ \\ $-1$ \\ $-1$ \end{tabular} & 
\begin{tabular}{c} $A^{({\rm o})}_{2g}$ \\ $E^{({\rm o})}_{g}$ \\ $A^{({\rm o})}_{1u}$ \\ $E^{({\rm o})}_u$ \end{tabular}& \\
\end{tabular}
\end{ruledtabular}
\label{table_D3d}
\end{table}

In Table~\ref{table_D3d}, we show the possible pair amplitudes in the $s$-wave channel, $\hat{F}({\bm k},i\varepsilon_n)$. As mentioned in Sec.~\ref{sec:bulkOF}, $\eta_P$ is equivalent to $\eta_{\rm mom}$ in single-band centrosymmetric SCs. In multiorbital systems, however, $\eta_P$ provides different indicators to classify the gap symmetry and topology. As shown in Table \ref{table_D3d}, the class of $\eta _{\rm orb}=+1$ (ESEE/OTEE) takes any one of the orbital bases, $\sigma_0$, $\sigma _x$, and $\sigma_z$. The pairs with $\sigma_0$ and $\sigma_x$ are categorized to even parity ($\eta_P=+1$) pairing, while the pair with $\sigma_z$ is an odd parity ($\eta_P=-1$) pairing. The class with orbital parity $\eta _{\rm orb}=-1$ is uniquely determined by $\sigma_y$, where only the inversion odd parity pairing $\eta _{P}=-1$ is possible. Odd-frequency pairs with the multiband, multiorbital, and sublattice degrees of freedom have been investigated widely in multiband SCs and SC/topological insulator hybrid systems~\cite{blackshafferPRB13,komendovaPRB15,asanoPRB15,triola,tamura20,ebisu16,triolaPRB19,blackpRB13,komPRB15,schmidt,asanoNJP18,blackPRB12,keiPRB18,brePRL18,crePRB15,kasPRB17,cay17,cay22}.

In the same manner, the general form of the gap function is given by
\beq
\hat{\Delta} ({\bm k},{\bm R}) = i\sum^{n_{\Gamma}}_{j=1} \eta^{\Gamma}_j ({\bm R}) \hat{d}^{\Gamma}_j({\bm k})s_y.
\label{eq:delta_orbital}
\eeq
The pair potential must satisfy $\hat{\Delta}({\bm k},{\bm R})  = - \hat{\Delta}^{\rm tr}(-{\bm k},{\bm R}) $,
As listed in Table~\ref{table_D3d}, all on-site pairing functions are categorized into $\Gamma=A^{({\rm e})}_{1g}$, $A^{({\rm e})}_{1u}$, $A^{({\rm e})}_{2u}$, and $E_u^{({\rm e})}$, where their basis functions are $\hat{d}^{A_{1g}} = (\sigma_0, \sigma_x)$, $\hat{d}^{A_{1u}} = s_z\sigma_y$, $\hat{d}^{A_{2u}} =\sigma_z$, and $(\hat{d}^{E_{u}}_{1}, \hat{d}^{E_u}_2) = (s_x\sigma_y,s_y\sigma_y)$, respectively. In centrosymmetric materials, the inversion symmetry in Eq.~\eqref{eq:inv} requires the pair potential to obey 
${P}\hat{\Delta}({\bm k}) {P}^{\dag}=\eta_P \hat{\Delta}(-{\bm k}) $.
Sufficient criterions for realizing time-reversal-invariant topological superconductivity in centrosymmetric materials are that the parity of the pairing is odd ($\eta _P=-1$) and the number of Fermi surfaces enclosing time-reversal-invariant momenta ($\Gamma$ and $Z$ points) is odd~\cite{satoPRB10,fuPRL10}. The $A^{({\rm e})}_{1u}$, $A^{({\rm e})}_{2u}$, and $E^{({\rm e})}_u$ pairing states satisfy the condition when the Fermi surface encircles the $\Gamma$ point.

In this paper, we focus on the $E_u$ superconducting state. The pair potential is given by
\beq
\hat{\Delta}({\bm k})= \left(\eta_1 \hat{d}^{E_u}_1({\bm k}) + \eta_2 \hat{d}^{E_u}_2({\bm k})\right)is_y,
\label{eq:nematicOP}
\eeq
As the time-reversal symmetry imposes $\eta_{1,2}\in \mathbb{R}$, the $E_u$ nematic state is given by
\beq
(\eta_1,\eta_2) = \Delta_0(\hat{n}_x,\hat{n}_y).
\label{eq:nematic_eta}
\eeq
The uniaxial vector, $\hat{\bm n}=(\cos\phi_{\bm n},\sin\phi_{\bm n},0)$, is a subsidiary nematic order~\cite{fuPRB14}. This state has gap anisotropy in the low-lying quasiparticle excitation, and the direction of the gap minimum or point nodes is represented by the nematic angle as $\phi _{\rm node} = \phi _{\bm n}\pm \pi/2$ [see Fig.~\ref{fig:fs}(a)]. The spontaneous breaking of the rotational symmetry of the crystal ensures the quasi-Nambu Goldstone mode associated with the fluctuation of the nematicity angle~\cite{uematsu}.
The hexagonal warping in Eq.~\eqref{eq:warp} induces a three-fold symmetric potential to pin the order parameter $\hat{\bm n}$. The mirror operator in Eq.~\eqref{eq:mirror} acts on the nematic gap function as
\beq
M\hat{\Delta}({\bm k})M^{\rm tr} = \eta^{M}_{C} \hat{\Delta} (\underline{\bm k}), 
\label{eq:mirrord}
\eeq
where $\eta^{M}_{C}=+1$ for $\hat{\bm n}\parallel\hat{\bm o}$ and $\eta^{M}_{C}=-1$ for $\hat{\bm n}\perp\hat{\bm o}$. 
In accordance with the topological Blount's theorem~\cite{kobayashiPRB14}, the parity $\eta^{M}_{C}$ is important for the topological stability of point nodes in centrosymmetric time-reversal-invariant SCs. The point nodes are topologically protected only when $\hat{\bm n}$ is pinned to the direction of one of three mirror axes, i.e., $\hat{\bm n}=\hat{\bm o}$ or $\phi _{\bm n}=0$, $\pm \pi /3$, and $\pm 2\pi/3$ [see Fig.~\ref{fig:fs}(a)]~\cite{kobayashiPRB14}. Otherwise, the point nodes in the $E_u$ state are gapped out by the hexagonal warping term.

Another pairing in the $E_u$ representation is the chiral state with broken time-reversal symmetry, which is given by
\beq
(\eta_1,\eta_2) = \Delta_0(1, \pm i).
\label{eq:chiral_eta}
\eeq
By projecting the quasiparticle states onto the conduction band of the normal electrons, the $4\times 4$ matrix of the odd-parity gap function is mapped onto the ${\bm d}$-vector as $\hat{\Delta} \mapsto i{\bm s}\cdot{\bm d}({\bm k})s_y$, where the ${\bm d}$-vector is the $2\times 2$ matrix in the band representation. In the lowest order on ${\bm k}$, the ${\bm d}$-vector in the conduction band is given as 
\beq
{\bm d}({\bm k})=[\eta_1 v_zf_z,\eta_2 v_zf_z,v(\eta_1 k_x+\eta_2 k_y)]/|m_0|.
\label{eq:dchiral}
\eeq
Thus the chiral state is the non-unitary state with ${\bm d}\times {\bm d}^{\ast} \neq {\bm 0}$, and the Cooper pairs has nonvanishing magnetic moment, ${\bm m}\propto i\langle {\bm d}\times {\bm d}^{\ast}\rangle$~\cite{sig91,takagi22}, where $\langle \cdots \rangle$ is the Fermi surface average. The spin polarization splits the quasiparticles bands into $E_{\pm}({\bm k})$, leading to the two distinct superconducting gaps. The bulk excitation gaps at the Fermi momentum ${\bm k}_{\rm F}$ are represented by 
$ E^{\rm chiral}_{\pm}({\bm k}_{\rm F})= \sqrt{|{\bm d}({\bm k}_{\rm F})|^2 \pm |{\bm d}({\bm k}_{\rm F})\times {\bm d}^{\ast}({\bm k}_{\rm F})|}$.
The $E^{\rm chiral}_+$ band is fully gapped on the whole Fermi surface, while the $E^{\rm chiral}_-$ band becomes gapless at $k_x=k_y = 0$. 
The pairwise point nodes at $k_x=k_y = 0$ are regarded as the Weyl points with opposite monopole charge $\pm 1$, and the quasiparticles residing at the nodes behave as Weyl fermions. The ${\bm d}$-vector in Eq.~\eqref{eq:dchiral} can be regarded as the hybrid structure of the spin-polarized polar pair [$f_z({\bm k})\ket{\uparrow\uparrow}$] and the chiral pair $(k_x+ik_y)\ket{\uparrow\downarrow+\downarrow\uparrow}$. The chiral state is the prototype of Weyl SCs. This state is thus the prototype of Weyl SCs~\cite{Sato-Fujimoto,mizushimaJPSJ16}, and low-lying quasiparticles with the nontrivial Berry curvature are responsible for anomalous transport phenomena~\cite{kobayashiPRL18,ishiharaPRB19}. 

The stability of the nematic and chiral phases is discussed in Appendix.~\ref{sec:phase}. It is demonstrated that the change of the carrier density (chemical potential) and the hexagonal warping effect drives the nematic-to-chiral phase transition (see Fig.~\ref{fig:phase}). The superconducting gap of the chiral state becomes fully isotropic when the Fermi surface is an open cylindrical shape. The stability of the chiral state is attributed to the gain of the condensation energy due to the Fermi surface evolution.

\begin{table*}[t!]
\caption{Emergent Cooper pair amplitudes in each irreducible representation, where ``primary'' and ``orbital'' denote the symmetry classes of $\hat{\Delta}$ and the Cooper pair amplitudes induced by the orbital hybridization term [$m({\bm k})\sigma$ in Eq.~\eqref{eq:hti}], respectively. ``SOC $(v_z)$'' and ``SOC $(v)$'' are spin-orbit-coupling induced pair amplitudes, and ``Warping'' corresponds to the hexagonal warping term.
}
\begin{ruledtabular}
\begin{tabular}{cccccccc}
& $\Delta$ &  primary & ``orbital'' & SOC $(v_z)$ & SOC $(v)$ & Warping ($\lambda$) & \\
\hline
& $A^{({\rm e})}_{1g}$  &  ESEE & ESEE & ESOO/OSOE & ETOE/OTOO & ETOE/OTOO & \\
& $A^{({\rm e})}_{1u}$  &  ETEO & OTEE & ETOE & ETOE & OSOE &\\
& $A^{({\rm e})}_{2u}$  &  ESEE & OSEO & OSOE & ETOE & ETOE &\\
& $E^{({\rm e})}_{u}$  &  ETEO & OTEE & ETOE & ETOE,OSOE & ETOE &
\end{tabular}
\end{ruledtabular}
\label{table4}
\end{table*}

We now classify Cooper pair amplitudes for all irreducible representations ($A^{({\rm e})}_{1g}$, $A^{({\rm e})}_{1u}$, $A^{({\rm e})}_{2u}$, $E^{({\rm e})}_{u}$) in the bulk STIs. The pair potentials except for $A^{({\rm e})}_{1g}$ ($\sigma_0$) violate the condition in Eq.~\eqref{eq:odd_condition}, i.e., $\hat{h} \hat{\Delta} -\hat{\Delta}\hat{h}^{\ast}\neq 0$, and are accompanied by odd-frequency pair amplitudes in the bulk. 
In Table~\ref{table4}, we summarize Cooper pair amplitudes induced by the insulating gap ($m_0$, $m_1$, $m_2$), the spin-orbit coupling ($v$ and $v_z$), and the hexagonal warping ($\lambda$). Let us focus on pair amplitudes in $E_u$ nematic states with Eq.~\eqref{eq:nematic_eta}. Equation~\eqref{eq:Fodd3} indicates that the commutation relation, $[\hat{h}({\bm k}), \hat{\tilde{\Delta}}({\bm k}) ]$, determines the emergence of the odd-frequency pairing. 
Equation~\eqref{eq:Fodd2} is recast into 
\beq
\hat{\mathcal{F}}^{\rm odd}({\bm k},i\varepsilon_n) = -i\varepsilon_{n}
A({\bm k},i\varepsilon_n)
\left[\hat{h}({\bm k}), \hat{\tilde{\Delta}}({\bm k}) \right],
\label{eq:Fodd3}
\eeq
where $A({\bm k},i\varepsilon_n)$ is a scalar coefficient obeying $A({\bm k},i\varepsilon_n)=A(-{\bm k},i\varepsilon_n)=A({\bm k},-i\varepsilon_n)$.
In the same manner, the even-frequency pairing in Eq.~\eqref{eq:evenF} is
\begin{align}
\hat{\mathcal{F}}^{\rm even}({\bm k},i\varepsilon_n) = & C_1({\bm k},i\varepsilon_n)\hat{\tilde{\Delta}}({\bm k})
+C_2({\bm k},i\varepsilon_n)\left\{ \hat{h}({\bm k}), \hat{\tilde{\Delta}}({\bm k})\right\} \nn \\
&+C_3({\bm k},i\varepsilon_n) \hat{h}^{\prime}({\bm k})\hat{\tilde{\Delta}}({\bm k})
\hat{h}^{\prime}({\bm k}),
\label{eq:Feven3}
\end{align}
where we have introduced $\hat{h}^{\prime}({\bm k})\equiv\hat{h}({\bm k})-c({\bm k})$, and the coefficients obey $C_j({\bm k},i\varepsilon_n)=C_j(-{\bm k},i\varepsilon_n)=C_j({\bm k},-i\varepsilon_n)$ ($j=1,2,3$).
It is seen from Eqs.~\eqref{eq:Fodd3} and \eqref{eq:Feven3} that the symmetry of $\mathcal{F}^{\rm odd}$ is governed by the commutation relation, $[\hat{h}({\bm k}), \hat{\tilde{\Delta}}]$, while the anticommutation relation $\{\hat{h}^{\prime}({\bm k}), \hat{\tilde{\Delta}}\}$ and the term $\hat{h}^{\prime}({\bm k})\hat{\tilde{\Delta}}\hat{h}^{\prime}({\bm k})$ induce even frequency pairing whose symmetry is different from that of $\hat{\Delta}$.
The insulating gap ($m_0$, $m_1$, $m_2$) and the spin-orbit coupling $(v)$ can induce the OTEE and OSOE pairing, $\hat{\mathcal{F}}^{\rm odd} = \hat{\mathcal{F}}^{\rm OTEE} + \hat{\mathcal{F}}^{\rm OSOE}$, where both pair amplitudes are classified into the $E_u^{({\rm o})}$ representation,
\begin{gather}
\hat{\mathcal{F}}^{\rm OTEE}({\bm k},i\varepsilon_n) = \varepsilon_n A({\bm k},i\varepsilon_n) m({\bm k}) 
\left(\eta_1 s_y - \eta_2 s_x\right) \sigma _z, \\
\hat{\mathcal{F}}^{\rm OSOE}({\bm k},i\varepsilon_n) = v \varepsilon_n A({\bm k},i\varepsilon_n) 
\left[\eta _1 f_y({\bm k})-\eta_2f_x({\bm k})\right]\sigma _x.
\end{gather}
In the vicinity of the $\Gamma$ point, the former (latter) is the intra-orbital spin-triplet $s$-wave (inter-orbital spin-singlet $p$-wave) pairing channel. In the same manner, from the leading order expansion in $v/|m_0|$, $v_z/|m_0|$, $\lambda/|m_0|$, the ETOE pairs also emerge where the spin-orbit coupling with $v$ and $v_z$ and the hexagonal warping induce the intra-orbital pairing $(\eta_1 f_y({\bm k})- \eta_2f_x({\bm k}))s_z$ and $f_z({\bm k}) (\eta_1 s_x - \eta_2 s_y)$, and the inter-orbital pairing $f_{3\lambda}({\bm k}) s_y \sigma _x$, respectively.

\begin{figure}[t]
\includegraphics[width=85mm]{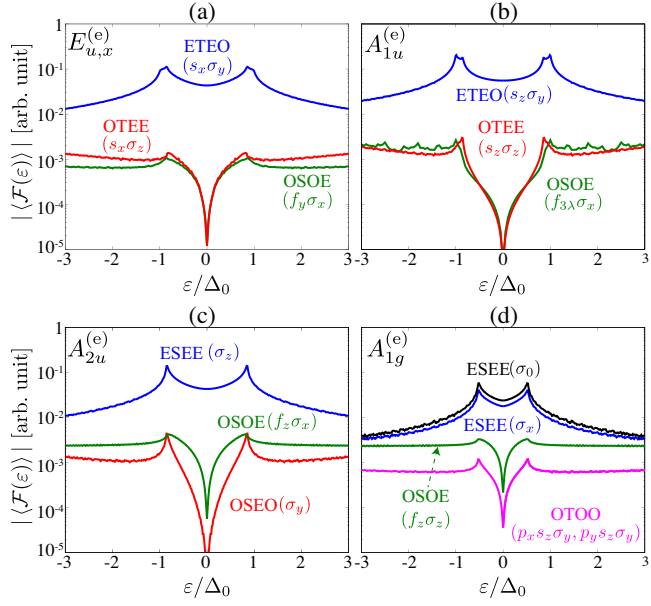}
\caption{Cooper pair amplitudes, $|\langle\mathcal{F}^{\rm ETEO}(\varepsilon)\rangle|$, $|\langle\mathcal{F}^{\rm OTEE}(\varepsilon)\rangle|$, and $|\langle\mathcal{F}^{\rm OSOE}(\varepsilon)\rangle|$, in $E^{({\rm e})}_{u,x}$ nematic state (a), the $A^{({\rm e})}_{1u}$ state (b), the $A^{({\rm e})}_{2u}$ state (c), and the inter-orbital ($\sigma_x$) $A^{({\rm e})}_{1g}$ state (d). We also plot the OSEO and OTOO pair amplitudes in (c) and (d), respectively. Here we consider bulk systems without surfaces, and set $\mu=1.65|m_0|$ and $\lambda k^2_{{\rm F},\parallel}/v= 0.1$ in all data. }
\label{fig:bulkOF}
\end{figure}

Figure \ref{fig:bulkOF} shows the Cooper pair amplitudes for $E^{({\rm e})}_{u}$ nematic ($s_x\sigma_x$), $A^{({\rm e})}_{1u}$, $A^{({\rm e})}_{2u}$, and $A^{({\rm e})}_{1g}$ ($\sigma_x$) superconducting states, where Cooper pair amplitudes at the energy $\varepsilon$ are obtained from the retarded/advanced Green's functions, $\check{G}^{\rm R,A}({\bm k},\varepsilon) = [\varepsilon \pm i\delta-\check{\mathcal{H}}({\bm k})]^{-1}$, with $\delta=0.01\Delta_0$. In the presence of odd-parity $E^{({\rm e})}_u$, $A^{({\rm e})}_{1u}$, and $A^{({\rm e})}_{2u}$ pair potentials, the dominant pair belongs to the ETEO class, which constitutes $\hat{\Delta}$ through the gap equation. As mentioned above, however, the orbital hybridization [$m({\bm k})\hat{\sigma}_x$] term and spin-orbit coupling terms induce OTEE and OSOE pair amplitudes in the odd-parity pairing states, though they are one or two order of magnitude smaller than ETEO pair amplitudes. In $A^{({\rm e})}_{2u}$ and $A^{({\rm e})}_{1g}$, the nonzero amplitudes of the OSEO and OTOO pairs also stem from the orbital hybridization and spin-orbit coupling with $v$, respectively.

\section{Odd-frequency pairs on surface of nematic and chiral states}
\label{sec:surface}

In this section, we discuss the surface ABSs and odd-frequency pairs in the $E_u$ nematic and chiral states. The quasiparticle states are described by the BdG equation
\begin{eqnarray}
\sum _{{\bm r}_j}\check{\mathcal{H}}({\bm r}_i,{\bm r}_j)
{\bm \varphi}_{n}({\bm r}_j)
= E_{n} {\bm \varphi}_{n}({\bm r}_i).
\label{eq:bdg}
\end{eqnarray}
The BdG Hamiltonian in real space is given by
\begin{align}
\check{\mathcal{H}}({\bm r}_i,{\bm r}_j) = 
\left(
\begin{array}{cc}
\hat{h}({\bm r}_i,{\bm r}_j) &
\hat{\Delta} ({\bm r}_i,{\bm r}_j) \\
\hat{\Delta}^{\dag}({\bm r}_i,{\bm r}_j) & 
- \hat{h}({\bm r}_i,{\bm r}_j)
\end{array}
\right),
\label{eq:Hbdg}
\end{align}
where $\hat{h}({\bm r}_i,{\bm r}_j)$ is the real-space coordinate representation of $\hat{h}({\bm k})$. 
The eigenvector is the eight-component vector, ${\bm \varphi}_{n}=[u_{n,(\uparrow,1)},u_{n,(\downarrow,1)},u_{n,(\uparrow,2)},u_{n,(\downarrow,2)},v_{n,(\uparrow,1)},v_{n,(\downarrow,1)},v_{n,(\uparrow,2)},v_{n,(\downarrow,2)}
]^{\rm tr}$, which fulfills the orthonormal condition,
$\sum _{{\bm r}_j} {\bm \varphi}^{\dag}_{n}({\bm r}_j){\bm \varphi}_{m}({\bm r}_j)  = \delta _{nm}$. 
The BdG Hamiltonian obeys the particle-hole symmetry, 
\beq
\mathcal{C}\check{\mathcal{H}}({\bm r}_i,{\bm r}_j)\mathcal{C}^{-1}=-\check{\mathcal{H}}({\bm r}_i,{\bm r}_j).
\label{eq:PHS}
\eeq
This guarantees that the
positive energy solution ${\bm \varphi}_{E}({\bm r})$ is
associated with the negative one ${\bm \varphi}_{-E}({\bm r}) = \mathcal{C}{\bm \varphi}_{E}({\bm r})$, where $\mathcal{C}=\tau _x K$ exchanges the particle and hole components of the Bogoliubov quasiparticle wavefunction. We numerically solve Eq.~\eqref{eq:bdg} with the periodic boundary condition in the $xy$ plane and the open boundary condition along the $z$-axis,
${\bm \varphi}_{k_x,k_y}(z=0) = {\bm \varphi}_{k_x,k_y}(z=L) ={\bm 0}$.
The $z$-axis corresponds to the (111) direction of the $D_{3d}$ crystals.

The nontrivial topology of the normal electrons in the parent material, Bi$_2$Se$_3$, is characterized by the $\mathbb{Z}_2$ invariant, that is, the parity of $m_0m_1$. When ${\rm sgn}(m_0m_1)<0$, the bulk of the parent material is topologically nontrivial and a gapless Dirac cone exists on the surface. It has been observed in angle resolved photoemission spectroscopy that the Dirac cone in Cu$_x$Bi$_2$Se$_3$ is well isolated from the bulk conduction band at the Fermi level, when the carrier density is small~\cite{wray}. The surface Dirac fermion is fully polarized in the orbital space and accompanied by helical spin texture in the surface Brillouin zone. It has been discussed that the isolated Dirac fermion significantly affects the surface ABSs and tunneling conductance~\cite{hsieh12,yamakage12,hao11,haoPRB17}, as well as the superconducting gap structure~\cite{mizushimaPRB14}.

\subsection{Nematic states}

\subsubsection{Chiral symmetry and topological invariant}

\begin{figure}[t!]
\includegraphics[width=85mm]{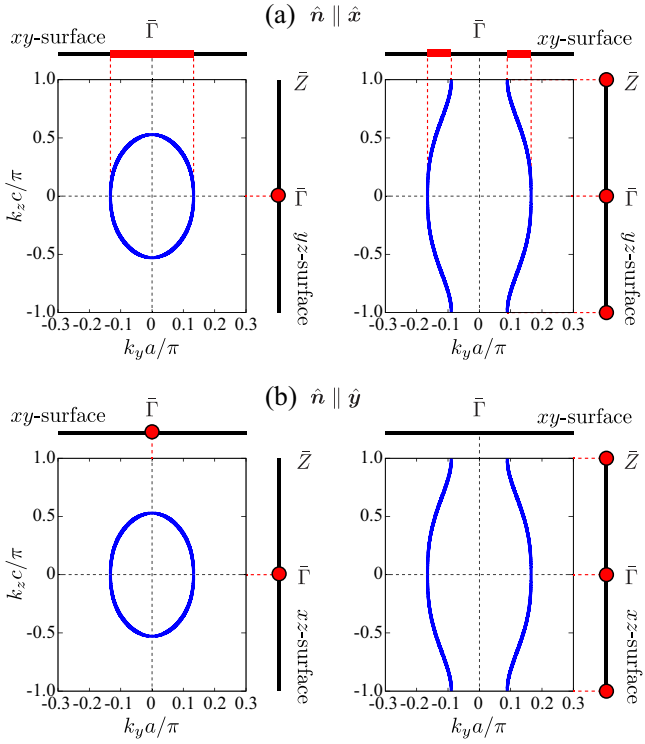}
\caption{Evolution of Fermi surfaces and the distribution of topologically protected zero energy states in the Brillouin zone projected on the $xy$ and $yz$ ($xz$) surfaces: (a) $E_{u,x}$ ($\hat{\bm n}\parallel\hat{\bm x}$) and (b) $E_{u,y}$ ($\hat{\bm n}\parallel\hat{\bm y}$) nematic states. The left and right figures correspond to $\mu = 1.6|m_0|$ and $1.9|m_0|$, respectively. In all data, we set $\lambda k^2_{{\rm F},\parallel}/v= 0.1$.}
\label{fig:topology}
\end{figure}

Figure \ref{fig:topology} shows the distribution of the zero energy states on surface Brillouin zones in the nodal $E_{u,x}$ state [$\hat{\bm n}=\hat{\bm x}$ in Eq.~\eqref{eq:nematic_eta}] and the fully gapped $E_{u,y}$ state [$\hat{\bm n}=\hat{\bm y}$ in Eq.~\eqref{eq:nematic_eta}]. The zero energy states are protected by two different topological invariants, $w_{\rm 3d}$ and $w_{\rm 1d}(k_y)$. The three-dimensional winding number is defined as $w_{\rm 3d}=\frac{1}{48\pi^3}\int d^3{\bm k}\epsilon^{\mu\nu\eta}{\rm tr}
[\check{\Gamma}_3 \check{Q}_{\mu}({\bm k})\check{Q}_{\nu}({\bm k})\check{Q}_{\eta}({\bm k})
]$ with $\check{Q}_{\mu}({\bm k})=\check{\mathcal{H}}^{-1}({\bm k})\partial _{k_{\mu}}\check{\mathcal{H}}({\bm k})$, where the repeated Greek indices imply the sum over $x,y,z$. The chiral operator is defined as a combination of the time-reversal symmetry in Eq.~\eqref{eq:trs} and particle-hole symmetry in Eq.~\eqref{eq:PHS}, $\check{\Gamma}_3=i\mathcal{C}\mathcal{T}$.
The winding number is nonzero for odd parity SCs when the Fermi surface encloses time reversal invariant momenta in the Brillouin zone. This ensures the existence of the zero energy states at the $\bar{\Gamma}$ point which is the center of the surface Brillouin zone, when an odd number of the Fermi surface encloses the $\Gamma$ point in the Brillouin zone. The parity of $w_{\rm 3d}$ still remains as a topological invariant even when the gap has point nodes~\cite{satoPRB10,fuPRL10,sasakiPRL11}.

Another topological invariant is the one-dimensional winding number, which is obtained from the order-two magnetic symmetry~\cite{satoPRB2009,mizushimaPRL12,mizushimaNJP13,tsutsumiJPSJ13},
\beq
\mathcal{T}\mathcal{M}\check{\mathcal{H}}(k_x,k_y,k_z) (\mathcal{T}\mathcal{M})^{-1} = \check{\mathcal{H}}(k_x,-k_y,-k_z),
\label{eq:P2symmetry}
\eeq
which is a combination of $\mathcal{T}$ and mirror reflection, $\mathcal{M}={\rm diag}(M,\eta^{M}_{C} M^{\ast})$. The mirror reflection operator on the $yz$ plane is defined as $M=-is_x$ and $\eta^{M}_{C}$ is given in Eq.~\eqref{eq:mirrord}. 
Combining $\mathcal{T}\mathcal{M}$ with Eq.~\eqref{eq:PHS},
we define the chiral symmetry operator for the $E_{u,x}$ nematic state ($\hat{\bm n}\parallel\hat{\bm x}$) as 
\beq
\check{\Gamma} _1  =e^{i\alpha} \mathcal{CTM} = s_z\tau_y,
\label{eq:chiral} 
\eeq
which is anti-commutable with the BdG Hamiltonian, 
$\{ \check{\Gamma} _1, \check{\mathcal{H}} (0,k_y,k_z)\} = 0$. 
The phase factor in Eq.~\eqref{eq:chiral} is determined so as to satisfy $\check{\Gamma}^2=+1$.
In the chiral symmetry plane  within $k_x=0$ [see Fig.~\ref{fig:fs}(a)], the one-dimensional winding number is defined as~\cite{satoPRB2009,satoPRB11,mizushimaNJP13,mizushimaPRL12}
\begin{align}
w_{\rm 1d}(k_{y}) = - \frac{1}{4\pi i}\int^{+\pi/c}_{-\pi/c} d k_{z}{\rm tr}\left[ 
\check{\Gamma}_1 \check{Q}_z(0,k_y,k_z)
\right].
\label{eq:w1d}
\end{align}

In the $E_{u,x}$ nematic state, the winding number is nontrivial, $w_{\rm 1d}(k_y=0)=-2$, for $|k_y|\le k_{{\rm F},y}$ unless the Fermi surface is opened to the $\hat{\bm z}$ direction. This leads to the topological stability of zero-energy flat band on the $xy$ surface Brillouin zone as shown in Fig.~\ref{fig:topology}(a). At the critical carrier doping, $\mu _{\rm c}=1.8|m_0|$, the Fermi surface evolution from a closed spheroidal to an opened cylindrical shape changes $w_{\rm 1d}(k_y)$ around the $\bar{\Gamma}$ point. 
In contrast, in the $E_{u,y}$ nematic state, the zero energy state appears only at $\bar{\Gamma}$ for a closed Fermi surface, while it vanishes beyond the critical doping $\mu > \mu_{\rm c}$ [Fig.~\ref{fig:topology}(b)]. We note that the symmetry in Eq.~\eqref{eq:P2symmetry} is not broken by Zeeman fields perpendicular to the $x$-direction, and thus the zero-energy flat band in the $E_{u,x}$ state is tolerant to such magnetic fields. 

\subsubsection{Surface Andreev bound states}
\label{sec:SABSs}

\begin{figure}[t]
\includegraphics[width=85mm]{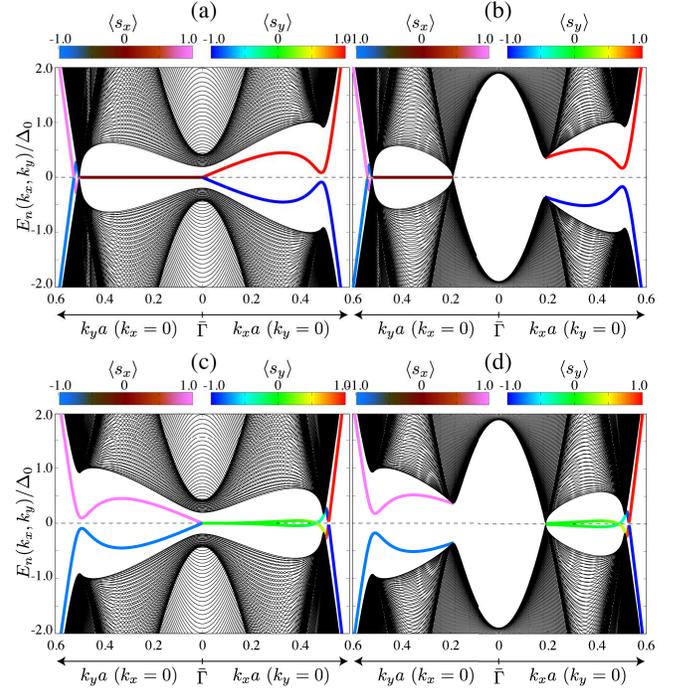}
\caption{Quasiparticle spectra on the $(111)$ surface of the $E_u$ nematic states: (a) $\mu/|m_0|=1.8$ and (b) $1.85$ for the $E_{u,x}$ nematic state ($\hat{\bm n}\parallel\hat{\bm x}$), and (c) $\mu/|m_0|=1.8$ and (d) $1.85$ for the $E_{u,y}$ nematic state ($\hat{\bm n}\parallel\hat{\bm y}$). Here we set $\lambda k^2_{{\rm F},\parallel}/v = 0.1$.}
\label{fig:Eunematic_z_spect}
\end{figure}

In Fig.~\ref{fig:Eunematic_z_spect}, we show the low-lying quasiparticle spectra on the $(111)$ surface of the $E_{u,x}$ and $E_{u,y}$ nematic states. As seen in Fig.~\ref{fig:Eunematic_z_spect}(a) and \ref{fig:Eunematic_z_spect}(b), in the $E_{u,x}$ state, the zero-energy flat band appears along $k_y$ as a consequence of nontrivial topological invariant in Eq.~\eqref{eq:w1d}. In the $E_{u,y}$ state, the zero energy state appears at the $\bar{\Gamma}$ point, which is dispersing to both $k_x$ and $k_y$ directions [Fig.~\ref{fig:Eunematic_z_spect}(c)]. The velocity is proportional to the strength of the hexagonal warping, $\lambda$, and the dispersive ABSs reduce to the flat band when $\lambda = 0$. 
The topological phase transition occurs at $\mu _{\rm c} = 1.8 |m_0|$, and a part of the zero energy flat band in the $E_{u,x}$ nematic state survives even when $\mu > \mu _{\rm c}$ [Fig.~\ref{fig:Eunematic_z_spect}(b)].
These results coincide with the topological consideration with $w_{\rm 3d}$ and $w_{\rm 1d}(k_y)$. Furthermore, the peculiarity of the surface ABSs in the STIs is that the gapless branch within $k_{x,y}a\lesssim 0.5$ smoothly evolves into the Dirac cone corresponding to the dispersion within $k_{x,y}a\gtrsim 0.5$. The Dirac cone always has the steep dispersion compared to the surface ABSs, because the dispersion of the former (latter) is governed by the insulating gap (superconducting gap).

In Fig.~\ref{fig:Eunematic_z_spect}, we also plot the spin polarization of the surface ABSs using the color bar. The spin-polarization rate of the quasiparticle state with $E_n({\bm k}_{\parallel})$ is defined as
\beq
\langle s_{\mu}({\bm k}_{\parallel})\rangle = \int^{L/2}_0 dz{\bm U}_{n}^{\dag}({\bm k}_{\parallel},z) 
{s}_{\mu}
{\bm U}_n({\bm k}_{\parallel},z) ,
\eeq
which is the expectation value of the spin operator by the particle-component of the quasiparticle wavefunction, ${\bm U}_n\equiv [u_{n,(\uparrow,1)},u_{n,(\downarrow,1)},u_{n,(\uparrow,2)},u_{n,(\downarrow,2)}]^{\rm tr}$. While the zero energy flat band does not possess characteristic spin texture, the dispersive surface ABSs, e.g., the $k_x$ ($k_y$) dispersion in $E_{u,x}$ ($E_{u,y}$), exhibit the helical spin texture $\langle {\bm s}\rangle \perp {\bm k}$. The spin texture of the surface ABSs smoothly evolves the helical spin texture of the surface Dirac cone. 

To capture the evolution of the surface ABSs, we compute the surface density of states (DOS). The local DOS is then introduced as $\mathcal{N}\equiv \sum _{s,\sigma}\mathcal{N}_{s,\sigma} $, where $\mathcal{N}_{s,\sigma}$ is defined as 
\begin{align}
\mathcal{N}_{s,\sigma}(z,\varepsilon) = - \frac{1}{\pi}\sum_{{\bm k}_{\parallel}} {\rm Im}
[\check{G}^{\rm R}_{{\bm k}_{\parallel}}(z,z,\varepsilon)]_{(s,\sigma),(s,\sigma)}.
\end{align}
The retarded Green's function, $\check{G}^{\rm R}$, is obtained from Eq.~\eqref{eq:G} with the analytic continuation, $\check{G}^{\rm R}(\varepsilon) = \check{G}(i\varepsilon_n \rightarrow \varepsilon + i\eta)$, where $\eta$ is an infinitesimal constant. For the computation of the surface Green's function, we utilize the recursive Green's function method using the prescription proposed by Umerski~\cite{umerski}.

\begin{figure}[t]
\includegraphics[width=85mm]{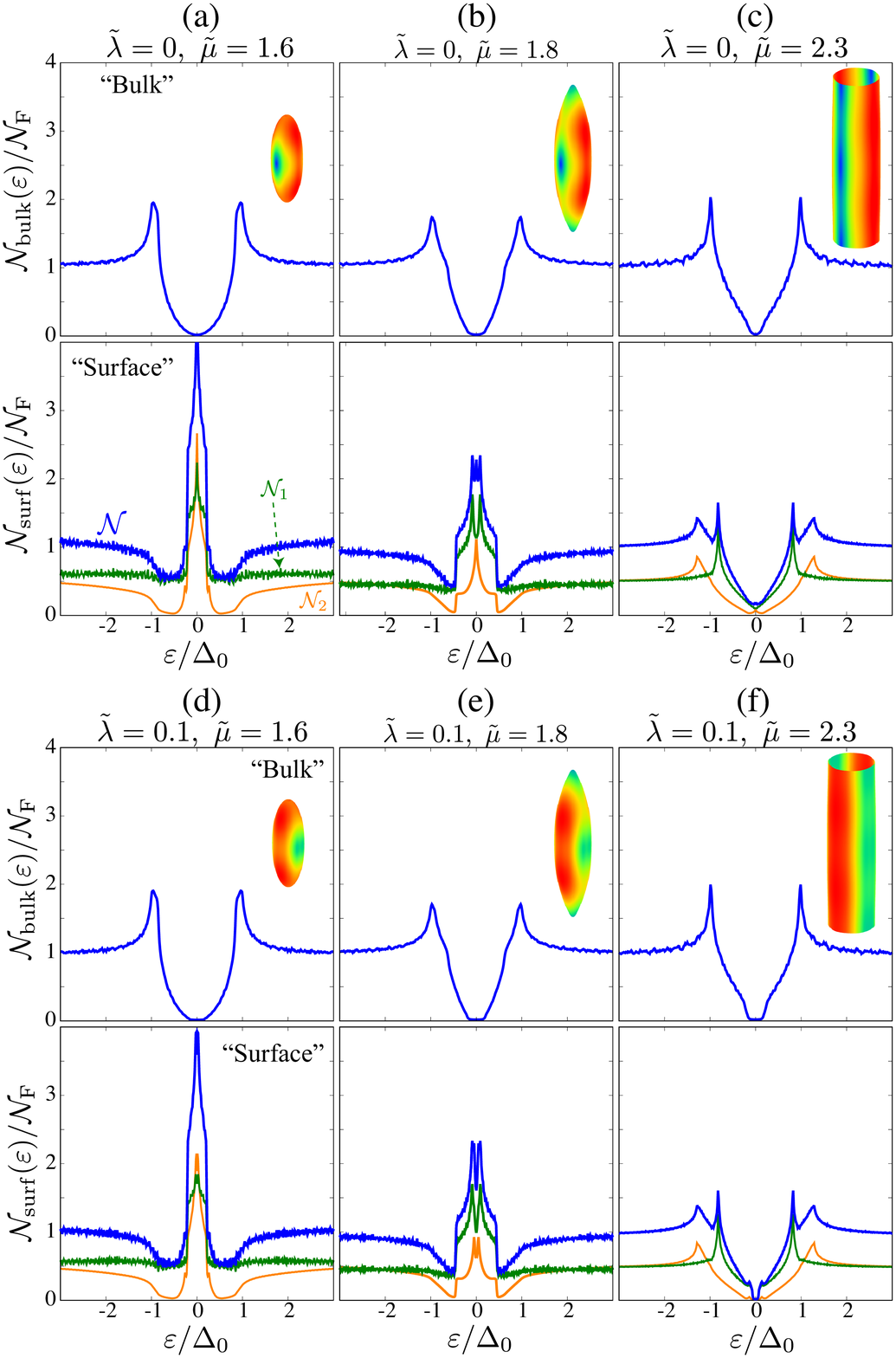}
\caption{Bulk and surface DOS, $\mathcal{N}(z=0,\varepsilon)$ in the $E_{u,x}$ nematic states (a-c) and $E_{u,y}$ nematic states (d-f): (a,d) $\mu/|m _0|=1.6$, (b,e) $\mu/|m _0|=1.8$, and (c,f) $\mu/|m _0|=2.3$, where we set $\lambda k^2_{{\rm F},\parallel}/v=0.1$. The inset figures show the Fermi surface at each $\mu$.}
\label{fig:nematic_sdos}
\end{figure}

Figure~\ref{fig:nematic_sdos} shows the bulk and surface DOS, $\mathcal{N}_{\rm bulk}(\varepsilon)\equiv\mathcal{N}(z=L/2,\varepsilon)$ and $\mathcal{N}_{\rm surf}(\varepsilon)\equiv\mathcal{N}(z=0, \varepsilon)$ in the $E_{u,x}$ and $E_{u,y}$ nematic states. The bulk superconducting gap in the $E_{u,x}$ state evolves from the point nodal to line nodal structure with increasing $\mu$. The bulk DOS in low energies changes from $\mathcal{N}\propto |\varepsilon|^2$ to the V-shape. Although the zero-energy flat band becomes dispersive in the $E_{u,y}$ state, the surface density of state exhibits the pronounced zero-energy peak irrespective of the nematic angle. 
The Fermi surface evolution gives rise to the topological phase transition and the zero energy state at $\bar{\Gamma}$ is gapped out. In the cylindrical Fermi surface, the characteristic peaks around $\varepsilon=0$ disappear and the low-energy surface DOS exhibits $\mathcal{N}(\varepsilon)\propto |\varepsilon|$. In Fig.~\ref{fig:nematic_sdos}, we also display the orbital-resolved DOS, $\mathcal{N}_{\sigma=1,2}(z,\varepsilon)\equiv \sum_{s=\uparrow,\downarrow}\mathcal{N}_{s,\sigma}(z,\varepsilon)$. The discrepancy between $\mathcal{N}_{1}$ and $\mathcal{N}_{2}$ at $\mu = 2.3|m_0|$ is attributed to the orbital polarization of the surface Dirac cones associated with the nontrivial topology of the normal electrons, which deviates the surface DOS from the bulk one. 

\subsubsection{Odd-frequency pairs and spectral bulk-boundary correspondence}

\begin{figure}[t]
\includegraphics[width=85mm]{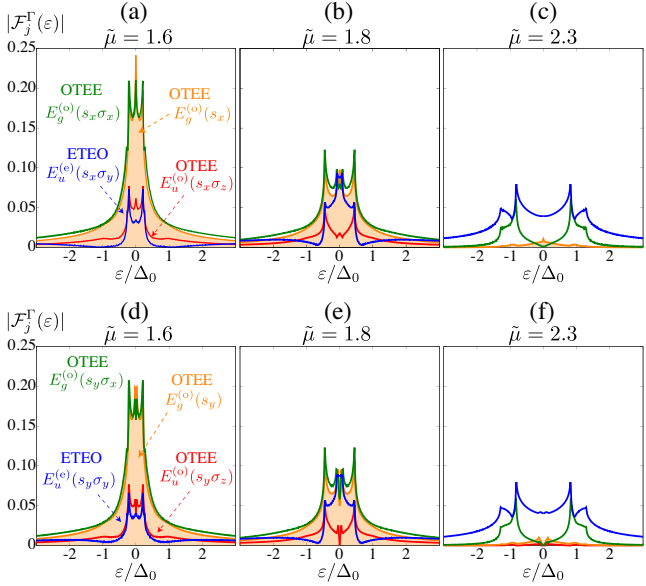}
\caption{Pair amplitudes on the surface, ${\mathcal{F}}^{\Gamma}_j (\varepsilon) \equiv \hat{\mathcal{F}}^{\Gamma}_j(z=L/2,\varepsilon)$, in (a-c) the $E_{u,x}$ nematic state and (d-f) the $E_{u,y}$ nematic state: $\tilde{\mu}\equiv\mu/|m_0|=1.6$ (a,d) , $1.8$ (b,e) and $2.3$ (c,f). In all data, we set $\lambda k^2_{{\rm F},\parallel}/v=0.1$.}
\label{fig:OFsurface_nematic}
\end{figure}

Let us now discuss the relation between low-lying quasiparticles and odd-frequency pairs. Here we restrict our attention to the on-site Cooper pair amplitudes. The $s$-wave (on-site) components of Cooper pair amplitudes are defined by the anomalous Green's function, $\hat{\mathcal{F}}\equiv -i\hat{F}s_y$, as 
\begin{align}
\hat{\mathcal{F}}(z,\varepsilon) 
\equiv& \sum _{{\bm k}_{\parallel}}\left[
\hat{\mathcal{F}}^{\rm R}_{{\bm k}_{\parallel}}(z,z,\varepsilon)
-\hat{\mathcal{F}}^{\rm A}_{{\bm k}_{\parallel}}(z,z,\varepsilon)\right].
\end{align}
In general, the $s$-wave pair amplitudes are immune to non-magnetic disorders and can penetrate into dirty metals, while the non $s$-wave components are sensitive to disorders and do not cause the proximity effect to dirty metals.
We expand the $s$-wave Cooper pair amplitudes in terms of the basis function of the irreducible representation, $\hat{d}^{\Gamma}_j$ as 
\begin{align}
\hat{\mathcal{F}}(z,\varepsilon) 
=& \sum_{\Gamma}\sum_{j=1}^{n_{\Gamma}} \mathcal{F}^{\Gamma}_j(z,\varepsilon)\hat{d}^{\Gamma}_j,
\end{align}
where the expression of $\hat{d}^{\Gamma}_j$ is summarized in Table~\ref{table_D3d}.

Figure~\ref{fig:OFsurface_nematic} shows the Cooper pair amplitudes emergent on the surface of the $E_{u,x}$ and $E_{u,y}$ nematic states. As discussed in Sec.~\ref{sec:bulk}, there exist three different types of Cooper pairs: ETEO, OTEE, and OSOE pairs. For $E_{u,x}$ ($E_{u,y}$), the ETEO pairs constitute the bulk pair potential $\hat{\Delta}=s_x\sigma_y$ ($\hat{\Delta}=s_y\sigma_y$), while the OTEE pair, $s_x\sigma_z$ ($s_y\sigma_z$), and OSOE pair $f_y\sigma_x$ ($f_x\sigma_x$), stem from the orbital hybridization term and spin-orbit coupling term of the normal state Hamiltonian, respectively. We find that the OTEE pairs emergent on the surface can be classified in terms of the irreducible representation as 
\begin{align}
\hat{\mathcal{F}}^{\rm OTEE}(z,\varepsilon) =  &
\mathcal{F}^{E^{({\rm o})}_g}_1(z,\varepsilon)s_x + \mathcal{F}^{E^{({\rm o})}_g}_2(z,\varepsilon)s_x\hat{\sigma}_x\nn \\
&+ \mathcal{F}^{E^{({\rm o})}_u}(z,\varepsilon)s_x\hat{\sigma}_z.
\label{eq:FOTEE}
\end{align}
The first two terms ($s_x$ and $s_x\sigma_x$) belonging to $E_g^{({\rm o})}$ are proper to the surface, while the third term $s_x\sigma_z$ is the odd-frequency pairs residing in the bulk. 

One of the OTEE pair amplitudes proper to the surface, $\mathcal{F}^{E^{({\rm o})}_g}_1s_x$, is directly associated with Majorana fermions through the SBBC. The SBBC is an extension of the bulk-boundary correspondence of the chiral symmetric systems to the complex frequency, and clarify the relation between the spectral singularity of the odd-frequency pair amplitudes and the existence of the Majorana fermions~\cite{tamuraPRB19}. The SBBC is explicitly written with the complex frequency $\omega \in \mathbb{C}$ as 
\beq
{F}^{\rm SBBC}(k_y,\omega) = \frac{W(k_y,\omega)}{\omega}.
\label{eq:sbbc1}
\eeq
The quantity in the left hand side is the Cooper pair amplitude integrated along the $z$ direction, defined with the chiral operator as
\begin{align}
{F}^{\rm SBBC}(k_y,\omega) 
= \sum_j {\rm tr}\left[\check{\Gamma}_1\check{G}_{k_x=0,k_y}(z_j,z_j,\omega)\right],
\end{align}
where the sum $\sum_j$ is taken over the semi-infinite system. The pair amplitude is an odd function of $\omega$, $\hat{F}^{\rm SBBC}(k_y,\omega)=-\hat{F}^{\rm SBBC}(k_y,-\omega)$. The right hand side of Eq.~\eqref{eq:sbbc1} is evaluated from the bulk Green's functions $\check{G}({\bm k},\omega)=[\omega-\check{\mathcal{H}}({\bm k})]^{-1}$ and the generalization of the one-dimensional winding number \eqref{eq:w1d} to the complex frequency plane, $\lim _{\omega\rightarrow 0}W(k_y,\omega) = w_{\rm 1d}(k_y)$. 
Therefore, the SBBC states that the bulk quantity $W(k_y,\omega)$ is related to the accumulation of the Cooper pair amplitudes at the boundary. 
The odd-frequency pair amplitude accumulated at the boundary is expressed in the low-frequency limit as 
\beq
F^{\rm SBBC}(k_y,\omega) = \frac{w_{\rm 1d}(k_y)}{\omega} + \chi(k_{y})\omega + O(\omega^3)
\eeq
As demonstrated in Appendix~\ref{sec:sbbc}, the bulk quantity $\chi(k_y)$ exhibits a power law divergence, $\chi(k_y)\sim |k_y-k_{\rm c}|^{-2}$, where $k_{\rm c}$ denotes an endpoint of the zero-energy flat band in the surface Brillouin zone.
Therefore, the odd-frequency pairs associated with the SBBC have two different singularities at $\omega \rightarrow 0$ and $k_y \rightarrow k_{\rm c}$. 

In the $E_{u,x}$ nematic state, the chiral operator is defined in Eq.~\eqref{eq:chiral} as $\check{\Gamma}_1 = s_x\check{\tau}_y$. It is then found that the $E^{({\rm o})}_g$ component of the odd-frequency pair amplitudes contains  the chiral-symmetry-protected odd-frequency pair, 
\begin{align}
\sum_{k_y}{F}^{\rm SBBC}(k_y,\omega) \subset \sum_j \mathcal{F}^{E^{({\rm o})}_{g,s_x}}_1(z_j,\omega).
\label{eq:OTEE_SBBC}
\end{align}
In Appendix~\ref{sec:sbbc}, we numerically confirm that the $E^{({\rm o})}_g$ component of the OTEE pairs accumulated at the boundary is equivalent to the spectral features of the bulk, and obeys the topological critical behaviors at $\omega \rightarrow 0$ and $k_y \rightarrow k_{\rm c}$.

In addition to the $E_g^{({\rm o})}$ pairs, the OTEE pairs in Eq.~\eqref{eq:FOTEE} include the $E^{({\rm o})}_u$ component. As in Fig.~\ref{fig:OFsurface_nematic}(a), the large amplitude of the $E^{({\rm o})}_u$ pair appears on the surface though it is negligible in the bulk. Its amplification on the surface is attributed to the interplay of the surface ABSs with the orbital polarization of the surface Dirac fermions. This resembles that the surface Dirac fermions induce the parity mixing of the pair potential near the surface~\cite{mizushimaPRB14}. Although the $E^{({\rm o})}_u$ component exists even in the bulk, as seen in Fig.~\ref{fig:bulkOF}(a), the pair amplitude is a few order of magnitudes smaller than that of the even-frequency $E_{u}^{({\rm e})}$ pairs in the bulk. As seen in Fig.~\ref{fig:OFsurface_nematic}(a), however, the amplitude of the $E^{({\rm o})}_u$ component on the surface becomes comparable to that of $E_{u}^{({\rm e})}$ pairs. As mentioned in Sec.~\ref{sec:SABSs}, the gapless surface ABSs in the nematic state smoothly evolve into the steep dispersion of surface Dirac fermions around the momenta $k_{x}a \approx 0.5$ and $k_{y}a \approx 0.5$. The wave function of the Dirac fermions in the normal state is fully polarized in the orbital space as a consequence of the inversion symmetry breaking on the surface. The orbital polarization of the Dirac fermions gives rise to the orbital polarization of the OTEE pair amplitudes on the surface as well as the surface ABSs with $k_{\parallel}\gtrsim k_{\rm F}$. Hence, the possible form of the orbitally polarized OTEE pairs is given by the linear combination of $s_x$ and $s_x\sigma_z$ as $\mathcal{F}^{E^{({\rm o})}_g}_1s_x + \mathcal{F}^{E^{({\rm o})}_u}s_x\hat{\sigma}_z$. The $E^{({\rm o})}_g$ component exhibits a singular behavior as $\mathcal{F}^{E^{({\rm o})}_g}_1(\omega \rightarrow 0) \propto 1/\omega$, which induces a large parity mixing of the $E^{({\rm o})}_u$ component with $E^{({\rm o})}_g$. As the strong enhancement of the $E^{({\rm o})}_u$ component on the surface stems from the orbital polarization of the surface ABSs,
the amplitude of the $E^{({\rm o})}_g$ component reduces with increasing $\mu$ and disappears at $\mu/|m_0| = 2.3$ as shown in Fig.~\ref{fig:OFsurface_nematic}(c), where the nematic state with an open cylindrical shape of the Fermi surface is fully gapped and accompanied by no surface ABSs. 

We plot in Figs.~\ref{fig:OFsurface_nematic}(d-f) the emergent Cooper pair amplitudes in the $E_{u,y}$ nematic state. The $E_{u,y}$ state spontaneously breaks the symmetry in Eq.~\eqref{eq:P2symmetry} and the zero energy ABSs for  ${\bm k}_{\parallel}\neq {\bm 0}$ are gapped out by the warping term with $\lambda k^2_{{\rm F},\parallel}/v=0.1$, the symmetry and amplitude of the emergent pairs are essentially same as those of the $E_{u,x}$ state. Hence, the evolution of the surface ABSs and emergent odd-frequency pairs reflect the topological phase transition driven by the change of carrier concentration, while they are less sensitive to the nematic angle $\hat{\bm n}$.

\subsection{Chiral states}

Let us now turn to the chiral state in the $E_u$ representation, whose pair potential is obtained from Eq.~\eqref{eq:chiral_eta} as 
\beq
\hat{\Delta} = \Delta _0 \left( s_x \sigma _y + i s_y \sigma_y \right).
\label{eq:delta_chiral}
\eeq 
For the weak coupling limit $\Delta_0 \ll \varepsilon_{\rm F}$, the gap structure is represented by the ${\bm d}$-vector in Eq.~\eqref{eq:dchiral}. As mentioned in Sec.~\ref{sec:bulk}, the ${\bm d}$-vector is the admixture of the chiral pair [$(k_x+ik_y)\ket{\uparrow\downarrow+\downarrow\uparrow}$] and the spin-polarized polar pair [$f_z(k_z)\ket{\uparrow\uparrow}$ with $f_z(k_z)=-f_z(-k_z)$], and the non-unitary chiral state has two distinct gaps in the bulk, $E^{\rm chiral}_{\pm}$. For a closed Fermi surface, the $E^{\rm chiral}_-$ band has pairwise Weyl points at $k_x=k_y=0$, which are responsible for the appearance of the zero-energy flat band of the spin-polarized chiral Majorana fermions on the $x$- and $y$-surface. For the (111) surface, no characteristic surface states stem from the Weyl point in the $E^{\rm chiral}_-$ band topology. 

\begin{figure}[t!]
\includegraphics[width=85mm]{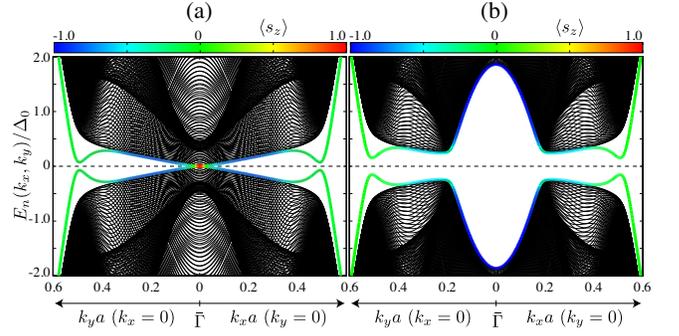}
\caption{Quasiparticle spectra on the $(111)$ surface of the $E_u$ chiral state: (a) $\mu/|m_0|=1.80$ and (b) $1.85$, where we set $\lambda = 0$.}
\label{fig:Eu_z_spect}
\end{figure}

Figure~\ref{fig:Eu_z_spect} shows the low-lying quasiparticle spectra on the $(111)$ surface of the $E_u$ chiral state. For a closed Fermi surface in Fig.~\ref{fig:Eu_z_spect}(a), the low-lying branch is dispersing from the zero energy, which is embedded in the continuum excitations from the Weyl points in $E^{\rm chiral}_-({\bm k})$. This gapless state, which is absent in the bulk, is the surface ABSs which smoothly evolve to the surface Dirac cone. The gapless dispersion stems from the polar-like component of the pair potential, $f_z(k_z)\ket{\uparrow\uparrow}$, which turns out to be protected by the chiral symmetry.
The chiral state spontaneously breaks both the time-reversal symmetry ($\mathcal{T}=-is_yK$) and mirror reflection symmetry (${M}=-is_x$). The time-reversal operation flips the chirality of the gap function, but the flip of the chirality is compensated by the mirror reflection $x\mapsto -x$. Hence, the chiral state is invariant under the combined symmetry 
\beq
\mathcal{T}M\hat{\Delta}(\mathcal{T}M)^{\rm tr} = \hat{\Delta}.
\eeq
This leads to the chiral symmetry in Eq.~\eqref{eq:chiral}, and the well-defined winding number in Eq.~\eqref{eq:w1d}. The chiral gap function in Eq.~\eqref{eq:delta_chiral} reduces to the spin-polarized polar pair, $k_z\ket{\uparrow\uparrow}$, at $k_x=k_y=0$. Therefore, the winding number is nontrivial, $|w_{\rm 1d}(k_y=0)|=1$, unless the Fermi surface is opened in the $k_z$ direction. The nontrivial topology protects the zero energy eigenstate in the $\uparrow$ spin sector at $k_x=k_y=0$, that is, the spin-polarized Majorana zero mode. The spin-polarization rate $\langle s_z \rangle$ is plotted with color map in Fig.~\ref{fig:Eu_z_spect}. We numerically confirm that the zero energy state is fully polarized $\langle s_z \rangle = +1$, which is consistent to the  topological consideration. As $\mu$ further increases, as shown in Fig.~\ref{fig:Eu_z_spect}(b), the zero energy state is gapped out by the Fermi surface evolution to an open cylindrical shape.

\begin{figure}[t!]
\includegraphics[width=85mm]{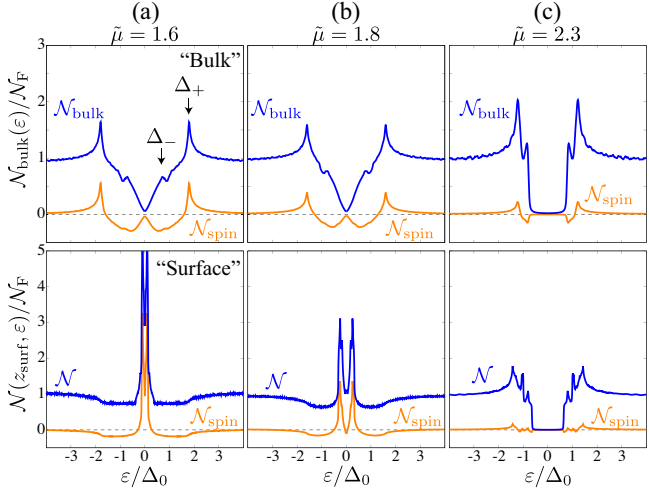}
\caption{Bulk DOS (upper row) and surface DOS (lower row) in the chiral state: (a) $\mu/|m_0|=1.6$, (b) $1.8$ and (c) $2.3$, where we set $\lambda = 0$. Here we also plot the spin-resolved DOS, $\mathcal{N}_{\rm spin}$.}
\label{fig:chiral_dos}
\end{figure}

We note that the spin-polarization of the surface bound states is different from that of the bulk quasiparticles. 
In Fig.~\ref{fig:chiral_dos}, we display the bulk and surface DOS in the chiral states for several $\mu$'s. The bulk DOS has two coherence peaks at $\varepsilon = \Delta_{+}$ and $\Delta_-$, corresponding to the two gaps of the non-unitary state, $\min E^{\rm chiral}_{+}$ and $\min E^{\rm chiral}_-$, respectively. In Fig.~\ref{fig:chiral_dos}, we also plot the spin-resolved DOS, $\mathcal{N}_{\rm spin}(z,\varepsilon)=\mathcal{N}_{\uparrow}(z,\varepsilon)-\mathcal{N}_{\downarrow}(z,\varepsilon)$, where $\mathcal{N}_{s}(z,\varepsilon)=\sum _{\sigma=1,2}\mathcal{N}_{s,\sigma}(z,\varepsilon)$. As mentioned above, in the bulk, the spin-polarization is induced by the splitting of quasiparticle bands due to non-zero spin density of Cooper pairs ${\bm m}\propto i{\bm d}({\bm k})\times {\bm d}^{\ast}({\bm k})$. 
According to the ${\bm d}$-vector representation of the chiral pair potential, the $(k_x+ik_y)\ket{\uparrow\downarrow+\downarrow\uparrow}$ component is accompanied by the spin-degenerate quasiparticles states around the Weyl points ($k_x=k_y=0$). In contrast, the polar component $f_z(k_z)\ket{\uparrow\uparrow}$ gap out the $\uparrow$-spin sector with the energy gap $\Delta_+$, leading to the gapless spectrum in the $\downarrow$-spin sector. This results in the sign change of the spin-resolved DOS at $\varepsilon = \pm \Delta_+$ in the bulk. The two coherence peaks approach each other and the spin-polarization rate decreases with increasing $\mu$. 

In contrast to the spin polarization of the bulk, as shown in Figs.~\ref{fig:chiral_dos}(a) and \ref{fig:chiral_dos}(b), the surface DOS in closed Fermi surfaces ($\mu = 1.6|m_0|$ and $1.8|m_0|$) has the sharp peaks around $\varepsilon=0$, which mainly stem from the gapless surface ABSs in the spin-$\uparrow$ sector. The intensity of the zero-energy peak is enhanced by the twist of the gapless surface ABSs around $k_xa \sim k_ya \sim 0.5$ [see Fig.~\ref{fig:Eu_z_spect}(a)]. Unlike the $E_{u,x}$ nematic state, only the zero energy state at the $\bar{\Gamma}$ point stems from nontrivial bulk topology and thus the zero energy peak is not protected by the topology. Hence, the two peaks of the surface DOS split and gradually spread as $\mu$ increases. The surface DOS becomes identical to the bulk one in the limit of the cylindrical Fermi surface [Fig.~\ref{fig:chiral_dos}(c)].

\begin{figure}[t]
\includegraphics[width=85mm]{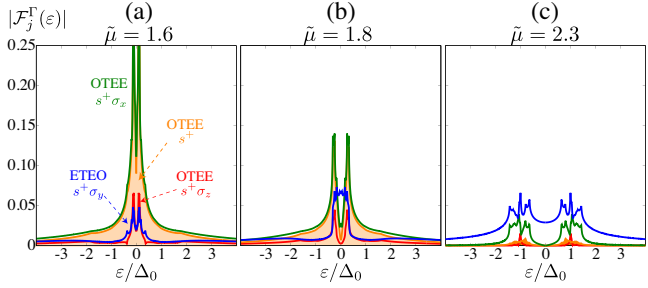}
\caption{Pair amplitudes on the surface of the chiral state: $\tilde{\mu}\equiv\mu/|m_0|=1.6$ (a), $1.8$ (b) and $2.3$ (c) with $\lambda=0$.}
\label{fig:OFsurface_chiral}
\end{figure}

In Fig.~\ref{fig:OFsurface_chiral}, we plot the emergent pair amplitudes on the surface of the chiral state for $\mu/|m_0|=1.6$, $1.8$ and $2.3$. Similarly with the nematic state, the OTEE pair amplitude is expanded in terms of the irreducible representation as 
\begin{align}
\hat{\mathcal{F}}^{\rm OTEE}(z,\varepsilon) =  &
\mathcal{F}^{E^{({\rm o})}_g}_1(z,\varepsilon)s^+ + \mathcal{F}^{E^{({\rm o})}_g}_2(z,\varepsilon)s^+\hat{\sigma}_x\nn \\
&+ \mathcal{F}^{E^{({\rm o})}_u}(z,\varepsilon)s^+\hat{\sigma}_z,
\label{eq:FOTEEchiral}
\end{align}
while the ETEO pair $s^+\sigma_y$ survives on the surface. Here $s^+\equiv s_x +is_y$ reflects the spin polarization of the chiral pair potential, giving rise to the spin-polarization of the surface ABSs and Cooper pair amplitudes. The emergent Cooper pairs are essentially the same as those in the nematic state, except for the spin-polarization. The $E^{({\rm o})}_g$ component of the OTEE pairs is responsible for the pronounced zero-energy peak of the surface DOS, and accompanied by the $E^{({\rm e})}_u$ component as a consequence of the parity mixing effect induced by surface Dirac fermions with strong orbital polarization. As $\mu$ increases through the topological phase transition, however, both the $E^{({\rm o})}_g$ and $E^{({\rm e})}_u$ components associated with the surface ABSs disappear.

\section{Odd-frequency pairs in junctions}
\label{sec:juntion}

In the previous sections, we have classified the odd-frequency Cooper pairs emergent in the bulk and surface. Here we discuss the contributions of odd-frequency pairs to the tunneling conductance and anomalous proximity effect in the STI junctions. 

\begin{figure}[t]
\includegraphics[width=85mm]{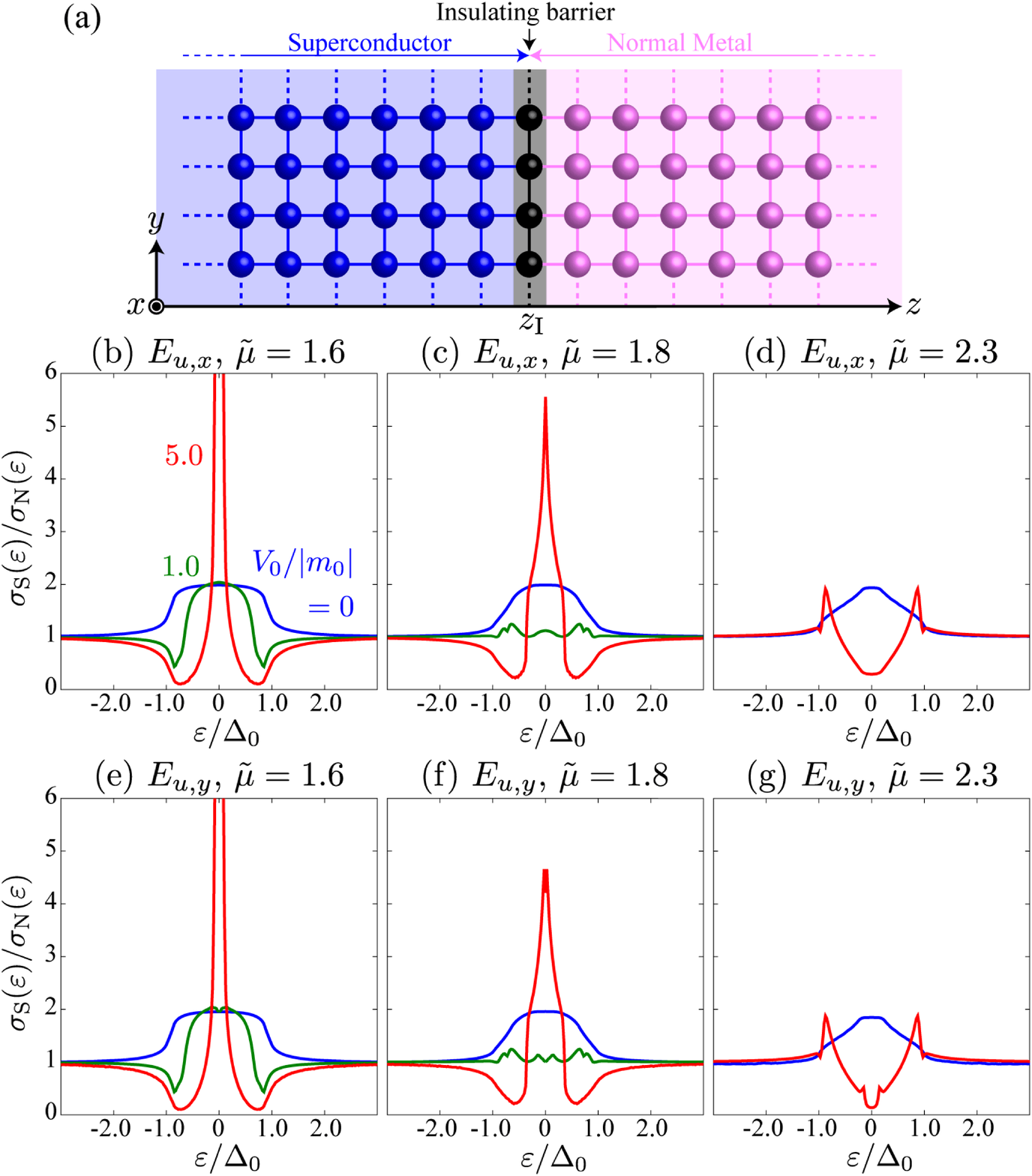}
\caption{(a) Schematics of a STI/normal metal junction, where the SC, the insulator, and the normal metal are stacked along the $z$ direction. (b-g) Normalized tunneling conductance $\sigma_{\rm S}(\varepsilon)/\sigma_{\rm N}(\varepsilon)$ in STI/normal metal junctions: (b-d) $E_{u,x}$ nematic state and (e-g) $E_{u,y}$ nematic state for several chemical potentials $\tilde{\mu}\equiv \mu/|m_0|=1.6$, $1.8$, and $2.3$. In all data, we set $\lambda k_{{\rm F},\parallel}^2/v=0.1$.}
\label{fig:SN}
\end{figure}

\subsection{Tunneling conductance in STI/norma metal junctions}
\label{sec:tunnel}

\begin{figure*}[t]
\includegraphics[width=140mm]{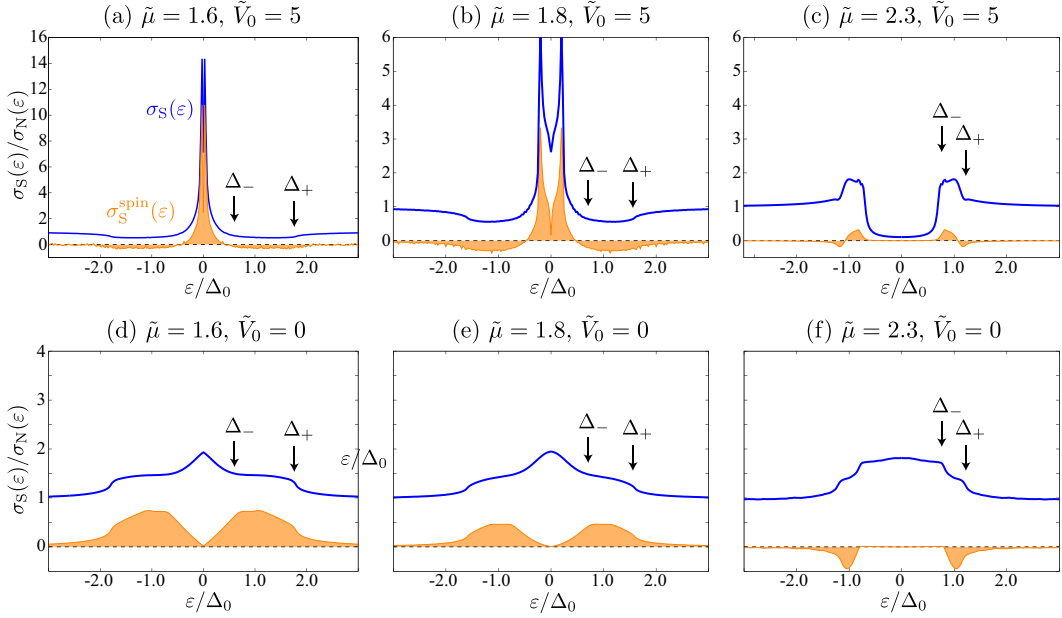}
\caption{Normalized tunneling conductance $\sigma_{\rm S}(\varepsilon)/\sigma_{\rm N}(\varepsilon)$ in the chiral state for the interface of low transmissivity (a-c) and high transmissivity (d-f): $\mu/|m_0|=1.6$ (a,d), $1.8$ (b,e), and $2.3$ (c,f). We also plot the spin-resolved tunneling conductance $\sigma^{\rm spin}_{\rm S}=\sigma_{\uparrow}-\sigma_{\downarrow}$. In all data, we set $\lambda = 0$.}
\label{fig:SN2}
\end{figure*}

Let us consider the tunneling conductance in a STI/normal metal junction. The tunneling spectroscopy has been established as a direct probe for the surface ABSs and the odd-frequency pairs, especially in the low transmissivity limit~\cite{tan95,kashiwaya00,tanaka12}. The junction which we consider here is depicted in Fig.~\ref{fig:SN}(a). The junction is stacked along the $z$-direction, i.e., the $(111)$ direction of the $D_{3d}$ crystal, and an insulating film is situated between the SC ($z<z_{\rm I}$) and the normal metal ($z>z_{\rm I}$). The translational symmetry is imposed in the $xy$ plane. Here we consider the Hamiltonian in Eq.~\eqref{eq:hti} for the normal metals.  Thus, the Hamiltonian in the normal state is given by adding the potential reflecting the insulating barrier at the interface, $V({\bm r})=V_0 \delta(z-z_{\rm I})$, as $\hat{h}({\bm r}_i,{\bm r}_j)\rightarrow \hat{h}({\bm r}_i,{\bm r}_j) + V({\bm r}_i)\delta_{ij}$, where $V_{\rm pot}$ determines the transmissivity of electrons at the interface $z_{\rm I}$. We also replace $\hat{\Delta}$ with $\hat{\Delta}(z_i)=\hat{\Delta}\Theta(z_{\rm I}-z_i)$, where $\Theta(x)$ is the step function.

In order to compute the tunneling conductance in STI/normal metal junctions, we derive the current operator in STIs. Here we consider an electric field applied to the $z$-direction,
$E_z({\bm r},t) = E_z({\bm r})e^{-i\omega t}$.
The gauge-invariance of the Hamiltonian requires the creation operator of an electron at a site to obey the gauge transformation $c^{\dag}_{\alpha}({\bm r})\rightarrow c^{\dag}_{\alpha}({\bm r})e^{-ie\Lambda({\bm r})}$, where $\Lambda({\bm r})$ is an arbitrary scalar function and the gauge field is transformed as ${\bm A}\rightarrow {\bm A}+{\bm \nabla}\Lambda$. As a result, the gauge fields are incorporated by the Peierls substitution, $(c_1,m_1,v_z) \rightarrow (c_1,m_1,v_z)e^{i\phi_{ij}}$, where the hopping terms along the $z$-direction acquire the Peierls phase,
$\phi _{ij}\equiv e\int^{{\bm r}_i}_{{\bm r}_j}{\bm A}({\bm r}^{\prime})d{\bm r}^{\prime} \approx {\bm A}({\bm r})\cdot({\bm r}_i-{\bm r}_j)$.
The electric bond current density from site ${\bm r}$ to ${\bm r}+{\bm \delta }_4$ is then obtained from Eq.~\eqref{eq:Htotal} as 
${\mathcal J}_z = - \delta H/\delta A_z 
=\frac{1}{2}{\bm \Psi}({\bm r}_{i+1})\check{\mathcal{J}}_z{\bm \Psi}({\bm r}_{i})+{\rm h.c.}$
[for the definition of ${\bm \delta}_4$ see Fig.~\ref{fig:fs}(a)], where the current operator in the particle-hole space is given by
\begin{align}
\check{\mathcal{J}}_z \equiv \begin{pmatrix}
\hat{j}_z & 0 \\ 0 & -\hat{j}_z^{\ast}
\end{pmatrix}, 
\end{align}
with 
\begin{align}
\hat{j}_z = -ie\left[-(c_1/c^2) \sigma _0
-(m_1/c^2) \sigma_x
+ i (v_z/c) \sigma_y/2\right].
\end{align}
The tunneling conductance in the STI/normal metal junction, $\sigma_{\rm S}(\varepsilon)$, is obtained from the Lee-Fisher formula in terms of the retarded and advanced Green's functions in the equilibrium, $G^{\rm R}$ and $G^{\rm A}$ as~\cite{leePRL81}
\begin{align}
\sigma_{\rm S}(\varepsilon) =&\frac{1}{2h}\sum _{{\bm k}_{\parallel}}
{\rm tr}^{\prime}\left[
\check{\mathcal{J}}_z\check{G}^{\prime\prime}_{z,z+1}\check{\mathcal{J}}_z\check{G}^{\prime\prime}_{z,z+1}
+\check{\mathcal{J}}^{\dag}_z\check{G}^{\prime\prime}_{z+1,z}\check{\mathcal{J}}^{\dag}_z\check{G}^{\prime\prime}_{z+1,z} \right. \nn \\
& \left. 
+\check{\mathcal{J}}_z\check{G}^{\prime\prime}_{z,z}\check{\mathcal{J}}^{\dag}_z\check{G}^{\prime\prime}_{z+1,z+1}
+\check{\mathcal{J}}^{\dag}_z\check{G}^{\prime\prime}_{z+1,z+1}\check{\mathcal{J}}_z\check{G}^{\prime\prime}_{z,z}
\right],
\end{align}
where we set $\check{G}^{\prime\prime}_{z,z^{\prime}}\equiv \check{G}^{\rm R}_{{\bm k}_{\parallel}}(z,z^{\prime},\varepsilon)-\check{G}^{\rm A}_{{\bm k}_{\parallel}}(z,z^{\prime},\varepsilon)$. The Green's function at the interface is numerically solved by using the recursive Green's function method combined with the M\"{o}bius transformation~\cite{umerski,yadaJPSJ14,takagi}.

In Fig.~\ref{fig:SN}, we plot the normalized tunneling conductance $\sigma_{\rm S}(\varepsilon)/\sigma_{\rm N}(\varepsilon)$ in the nematic states for different nematic angles $\hat{\bm n}=\hat{\bm x}$ and $\hat{\bm n}=\hat{\bm y}$, where $\sigma_{\rm N}$ is the tunneling conductance in the normal state. Here we compute $\sigma_{\rm S}/\sigma_{\rm N}$ for $V_0/|m_0|=0.0$, $1.0$, and $5.0$. The normalized conductance stays constant $\sigma_{\rm S}/\sigma_{\rm N}=2.0$ for $|\varepsilon|\le\Delta$ in high transmissivity limit ($V_0=0$), while a pronounced peak is developed at $\varepsilon=0$ with increasing $V_0$, i.e., with decreasing transmissivity. For $\hat{\bm n}=\hat{\bm x}$, as seen in Figs.~\ref{fig:SN}(b-d), the intensity of the ZBCP gradually weakens and the V-shaped gap is opened when $\mu$ increases across the topological phase transition. For $\hat{\bm n}=\hat{\bm y}$, where the mirror reflection symmetry is spontaneously broken by the nematic order, the tunneling conductance at $\mu/|m_0|=2.3$ has a subgap structure within $|\varepsilon|\sim 0.1\Delta$. This subgap structure reflects the uniaxially anisotropic full gap of the nematic state with $\hat{\bm n}=\hat{\bm y}$ due to the hexagonal warping effect. This is contrast to the case of $\hat{\bm n}=\hat{\bm y}$ where the nodal lines are protected by the mirror reflection symmetry. Hence, the spectral evolution of $\sigma_{\rm S}$ in the low transmissivity limit clearly captures the evolution of the surface DOS in Fig.~\ref{fig:nematic_sdos}.

Figure \ref{fig:SN2} shows the normalized tunneling conductance in the chiral state for the interface of low transmissivity ($V_0/|m_0|=5$) and high transmissivity ($V_0=0$). We also plot the spin-resolved tunneling conductance $\sigma^{\rm spin}_{\rm S}=\sigma_{\uparrow}-\sigma_{\downarrow}$. 
The spin-polarization of the odd-frequency pairs can be directly detected by scanning tunneling spectroscopy with ferromagnetic tip as a signature of the chiral state~\cite{takagi22}.

\subsection{Proximity effect to dirty normal metals}
\label{sec:proximity}

\begin{figure}[t!]
\includegraphics[width=85mm]{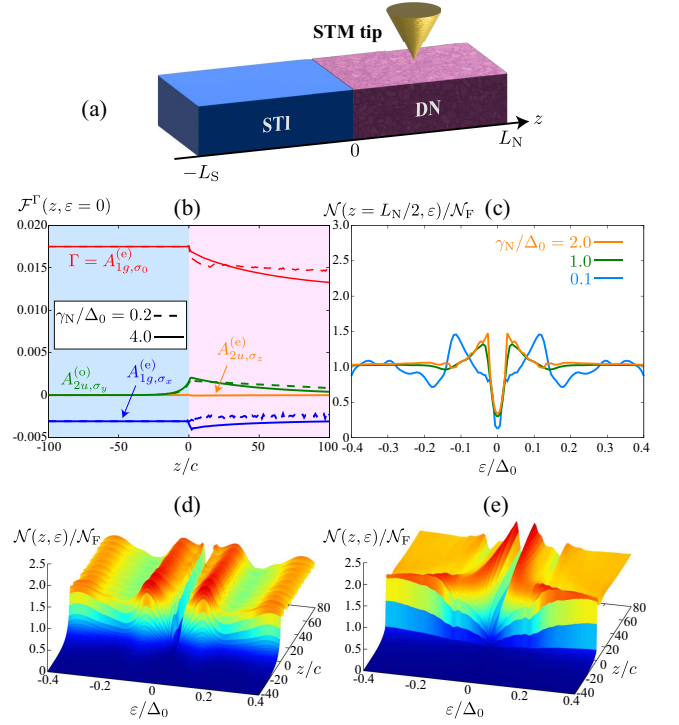}
\caption{(a) Schematics of a STI/DN junction, where the STI ($-L_{\rm S}<z\le 0$) is attached to the DN ($0<z<L_{\rm N}$). (b) Pair amplitudes in STI/DN junctions in the $A_{1g}^{({\rm e})}$ state with $\hat{\Delta}(z)=\Delta_0\sigma_0\Theta(-z)$. The DN in $z>0$ is the doped topological insulator with nonmagnetic impurities, $\gamma_{\rm N}/\Delta_0=0.2$ (dashed lines) and $1.0$ (solid lines). (c) Local DOS, $\mathcal{N}(z,\varepsilon)$, at $z=50c$ for $\gamma_{\rm N}/\Delta_0=0.1$, $0.2$, and $2.0$. The local DOS for $\gamma_{\rm N}/\Delta_0=0.2$ (d) and 4.0 (e). In all data, we set $\mu/|m_0|=1.6$ and $\lambda =0$.}
\label{fig:NS_F_A1g}
\end{figure}

Let us now discuss the anomalous proximity effect in the various irreducible representations of the STIs. We consider the STI/DN junction depicted in Fig.~\ref{fig:NS_F_A1g}(a), which is similar to that of Fig.~\ref{fig:SN}(a). The STI is attached to the DN without the insulating barrier at the interface. Let $L_{\rm S}$ and $L_{\rm N}$ be the thickness of the STI and DN region along the $z$-axis, respectively. The nonmagnetic impurities are absent in the STI region within $L_{\rm S}\le z \le 0$, while they are randomly distributed in the DN side within $0<z<L_{\rm N}$. In our calculation, we fix $L_{\rm S}/c=L_{\rm N}/c=150$.

The interaction of electrons with randomly distributed impurities is described by
\beq
\check{\mathcal{H}}_{\rm imp} = \sum^{N_{\rm imp}}_{a=1} \check{V}_{\rm imp} \delta({\bm r}-{\bm R}_a),
\eeq
where ${\bm R}_a$ is the position of an impurity atom and $N_{\rm imp}$ is the number of impurity atoms. The impurity potential, $\check{V}_{\rm imp}$, is the $8\times 8$ matrix in the spin, orbital, and particle-hole spaces. The full Green's function for $\check{\mathcal{H}}+\check{\mathcal{H}}_{\rm imp}$ is defined as 
$
\check{G}({\bm r}_i,{\bm r}_j,i\varepsilon_n)= [
\check{G}^{(0)-1}({\bm r}_i,{\bm r}_j,i\varepsilon_n) - \check{\mathcal{H}}_{\rm imp}({\bm r}_i)\delta({\bm r}_{ij})
]^{-1} 
$. The Green's function in the clean limit, $\check{G}^{(0)}$, is obtained from Eq.~\eqref{eq:G}. By taking the average over the randomly distributed impurities, the Green's function in equilibrium is governed by the Gor'kov equation in real space as~\cite{kopnin}
\begin{align}
\check{G}_{{\bm k}_{\parallel},ij} &= \check{G}^{(0)}_{{\bm k}_{\parallel},ij} 
+ \sum _{l} \check{G}^{(0)}_{{\bm k}_{\parallel},il} \check{\Sigma}_{{\rm imp},l}
\check{G}_{{\bm k}_{\parallel},lj},
\label{eq:NG}
\end{align}
where we introduce the abbreviations, $\check{G}_{{\bm k}_{\parallel},ij} \equiv \check{G}_{{\bm k}_{\parallel}}(z_i,z_j;i\varepsilon_n)$ and $\check{\Sigma}_{{\rm imp},l} \equiv \check{\Sigma}_{{\rm imp}}(z_l,i\varepsilon_n)$. Within the self-consistent Born approximation, the impurity self-energy $\check{\Sigma}_{\rm imp}$ is obtained with the renormalized Green's function as 
\beq
\check{\Sigma}_{\rm imp} (z,i\varepsilon _n) = n_{\rm imp}\sum _{{\bm k}_{\parallel}}
\check{V}_{\rm imp} \check{G}_{{\bm k}_{\parallel}}(z,z;i\varepsilon_n)\check{V}_{\rm imp}.
\label{eq:sigma_imp}
\eeq
where $n_{\rm imp}=N_{\rm imp}/V$ is the concentration of impurity atoms ($V$ is the volume of the system).

Here we consider the nonmangetic impurities independent of spin and orbital degrees of freedom, i.e., $\check{V}_{\rm imp} = v_{\rm imp}\check{\tau}_z$. Within the self-consistent Born approximation, the impurity scattering is characterized by the single parameter, that is, the scattering rate of the normal electrons, $\gamma_{\rm N}\equiv \pi \mathcal{N}_{\rm F}v^2_{\rm imp}n_{\rm imp}$. The impurity self-energy in Eq.~\eqref{eq:sigma_imp} is then recast into
\beq
\check{\Sigma}_{\rm imp} (z,i\varepsilon _n) = \Theta(z)
\frac{\gamma_{\rm N}}{\pi \mathcal{N}_{\rm F}}\sum _{{\bm k}_{\parallel}}
\check{\tau}_{z} \check{G}_{{\bm k}_{\parallel}}(z,z;i\varepsilon_n)\check{\tau}_{z}.
\label{eq:sigma_imp2}
\eeq
The Green's functions at equilibrium are determined by self-consistently solving Eqs.~\eqref{eq:NG} and \eqref{eq:sigma_imp2}. The retarded and advanced Green's functions are obtained by the analytic continuation of Matsubara Green's functions as $\check{G}^{\rm R}(\varepsilon) = \check{G}(i\varepsilon_n\rightarrow \varepsilon + i0_+)$ and $\check{G}^{\rm A}(\varepsilon) = \check{G}(i\varepsilon_n\rightarrow \varepsilon - i0_+)$, respectively. From now on, we fix the chemical potential to be $\mu = 1.6|m_0|$. This corresponds to the closed spheroidal Fermi surface around the $\Gamma$ point and all odd-parity ($E^{({\rm e})}_u$, $A^{({\rm e})}_{1u}$, and $A^{({\rm e})}_{2u}$) superconducting states are in the topological phase.

\begin{figure}[t]
\includegraphics[width=85mm]{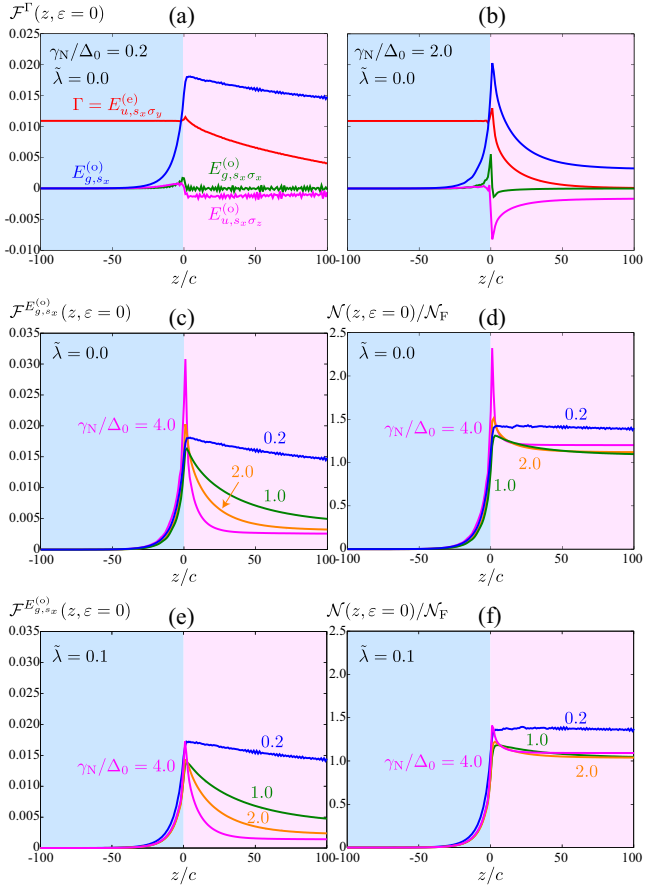}
\caption{(a,b) Pair amplitudes in STI/DN junctions for the $E_{u,x}^{({\rm e})}$ nematic state with $\hat{\Delta} =\Delta_0 s_x\hat{\sigma}_y\Theta(-z)$ and $\tilde{\lambda}\equiv\lambda k^2_{{\rm F},\parallel}/v= 0$. The nonmagnetic impurities in the DN region ($z>0$) are set to be $\gamma_{\rm N}/\Delta_0=0.2$ (a) and $2.0$ (b). (c,e) Zero-energy OTEE ($E_{g,s_x}^{({\rm o})}$) pair amplitudes and (d,f) local DOS at $\varepsilon=0$ for $\gamma_{\rm N}/\Delta_0=0.2$, 1.0, 2.0, and 4.0 at $\tilde{\lambda}= 0$ (c,d) and $0.1$ (e,f). In all data, we set $\mu/|m_0|=1.6$.
}
\label{fig:NS_F_nematic}
\end{figure}

Before going to the anomalous proximity effect in the odd-parity pairing states, we begin with the proximity effect in the conventional $s$-wave state. Figures~\ref{fig:NS_F_A1g}(b-d) show the Cooper pair amplitudes and the local DOS in the $A_{1g}^{({\rm e})}$, where the pair potential is given by $\hat{\Delta}=\Delta_0\hat{\sigma}_0$ for $z\le 0$ and $\hat{\Delta}=0$ for $z>0$. This is the conventional $s$-wave pairing state with a fully gap and accompanied by no low-lying ABS on the surface. In Fig.~\ref{fig:NS_F_A1g}(b), we plot the amplitude of the on-site Cooper pairs in the vicinity of the interface. The two $A_{1g}^{({\rm e})}$ components of the pair amplitudes ($\hat{\sigma}_0$ and $\hat{\sigma}_x$), which are dominant in the STI region, penetrate into the DN region. It is seen from Fig.~\ref{fig:NS_F_A1g}(b) that the proximitized $A_{1g}^{({\rm e})}$ pair amplitudes are insensitive to the increase of $\gamma_{\rm N}$ and thus survive in the DN region. As shown in Figs.~\ref{fig:NS_F_A1g}(c-e), the penetration of the even-parity pairs result in the opening of the finite energy gap in the DN region, where the gap is of the order of the Thouless energy depending on the diffusion constant and the length of the DN~\cite{gol88,bel96}.

The pair amplitudes in the $E_{u,x}^{({\rm e})}$ nematic state are displayed in Fig.~\ref{fig:NS_F_nematic}(a) $\gamma_{\rm N}/\Delta_0=0.2$ and \ref{fig:NS_F_nematic}(b) for $\gamma_{\rm N}/\Delta_0=2.0$, where we set $\lambda k^2_{{\rm F},\parallel}/v= 0.0$. The spatial profile of the pair potential is given by 
\beq
\hat{\Delta}(z) =\Delta_0 s_x\hat{\sigma}_y\Theta(-z).
\eeq
In the clean case ($\gamma_{\rm N}/\Delta_0=0.2$), the even-frequency odd-parity $E^{({\rm e})}_{u}$ pair, which constitutes the pair potential in the STI, penetrates into the DN, while the amplitude of the odd-frequency even-parity $E^{({\rm o})}_{g,s_x}$ pair significantly increases at the interface and penetrates into the DN region. It is seen from Fig.~\ref{fig:NS_F_nematic}(b) that as $\gamma_{\rm N}$ increases, the penetration depth of the odd parity $E^{({\rm e})}_u$ pair becomes short and its amplitude vanishes in $z/c\gtrsim 50$. The $\gamma_{\rm N}$-dependences of the odd-frequency even-parity ($E_{g,s_x}^{({\rm o})}$) pair amplitude and local DOS at $\varepsilon=0$ are plotted in Figs.~\ref{fig:NS_F_nematic}(c) and \ref{fig:NS_F_nematic}(d). In the clean case ($\gamma_{\rm N}/\Delta_0=0.2$), the $E_{g,s_x}^{({\rm o})}$ pairs penetrate into the DN region. The penetration of the odd-frequency pairs is responsible for the enhancement of the zero-energy peak of the local DOS in the DN region, $\mathcal{N}(z>0,\varepsilon=0)>\mathcal{N}_{\rm F}$, which is referred to as the anomalous proximity effect. As seen in Figs.~\ref{fig:NS_F_nematic}(c) and \ref{fig:NS_F_nematic}(d), however, the $E_{g,s_x}^{({\rm o})}$ pair amplitude is sensitive to the nonmagnetic impurity potential and suppressed in the DN region as $\gamma_{\rm N}$ increases, where the local DOS at $\varepsilon=0$ reduces to $\mathcal{N}(z>0,\varepsilon)\approx \mathcal{N}_{\rm F}$. Hence, the anomalous proximity effect in the nematic state of STIs is not immune to nonmagnetic impurities.

We note that as $\gamma_{\rm N}$ increases, the odd-frequency even-parity pair amplitude is tightly accumulated in the vicinity of the interface [see Figs.~\ref{fig:NS_F_nematic}(c)]. As shown in Fig.~\ref{fig:NS_F_nematic}(d), the localization of the $E_{g,s_x}^{({\rm o})}$ pair gives rise to the pronounced peak of the local DOS at $\varepsilon=0$. However, the peak structure of the pair amplitude and local DOS is not robust and depends on the detail of the parameters in the Hamiltonian. Indeed, it disappears when the hexagonal warping increases [Figs.~\ref{fig:NS_F_nematic}(e) and \ref{fig:NS_F_nematic}(f)].

In contrast to ordinary spin-triplet SCs, the anomalous proximity effect of the $E_u$ nematic state of the STI is fragile to nonmagnetic impurities. To understand the fragility of the anomalous proximity effect, we revisit the anomalous proximity effect in ordinary spin-triplet SCs which have no spin-orbit coupling effect. 
As an example, in Appendix~\ref{sec:dirac}, we demonstrate the anomalous proximity effect in Dirac SCs with the $d$-vector given by Eq.~\eqref{eq:dirac_d}. The superconducting state has a similar gap structure and surface ABSs with those of the $E_{u,x}$ nematic state [see Fig.~\ref{fig:helical_disp}(a)], except the absence of spin-orbit coupling; The pairwise point nodes appear on the $k_y$ axis in the bulk Brillouin zone and the zero-energy flat band on the surface is protected by the chiral symmetry. As shown in Fig.~\ref{fig:helical_dn}, the odd-frequency even-parity pairs penetrating into the DN region are insensitive to nonmagnetic impurities, which are responsible for the pronounced zero-energy peak of the local DOS in the DN region. The zero energy peak is immune to nonmagnetic impurities, when spin-orbit coupling is absent~\cite{tanPRB04}. 

\begin{figure}[t]
\includegraphics[width=85mm]{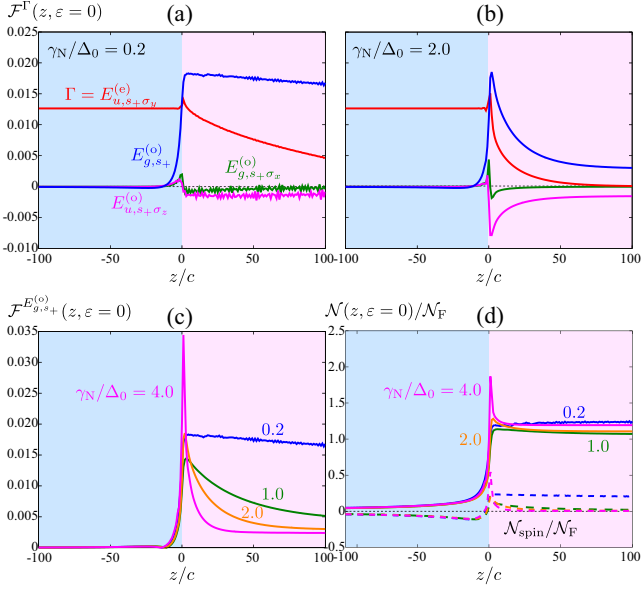}
\caption{(a,b) Pair amplitudes in STI/DN junctions for the chiral state. The nonmagnetic impurities in the DN region ($z>0$) are set to be $\gamma_{\rm N}/\Delta_0=0.2$ (a) and $2.0$ (b). (c) Zero-energy OTEE ($E_{g,s_+}^{({\rm o})}$) pair amplitudes and (d) local DOS at $\varepsilon=0$ for $\gamma_{\rm N}/\Delta_0=0.2$, 1.0, 2.0, and 4.0. The OTEE pairs are not protected by the chiral symmetry. In all data, we set $\mu/|m_0|=1.6$ and $\lambda k^2_{{\rm F},\parallel}/v= 0$.
}
\label{fig:NS_F_chiral}
\end{figure}

The $\gamma_{\rm N}$-dependence of the anomalous proximity effect in the nematic state of STIs is essentially different from that in Dirac SCs. We attribute this discrepancy to strong spin-orbit coupling inherent to the parent material of the STIs. In the $E_{u,x}$ nematic state, as shown in Fig.~\ref{fig:NS_F_nematic}, the dominant component of proximitized pairs belong to the $E^{({\rm o})}_g$ irreducible representation with basis function, $s_x\hat{\sigma}_0$. Let $\hat{U}\in U(4)$ be a unitary matrix that diagonzalizes the normal-state Hamiltonian $\hat{h}({\bm k})$ as $\hat{U}^{\dag}\hat{h}({\bm k})\hat{U} = {\rm diag}(\varepsilon_+({\bm k}),\varepsilon_+({\bm k}),\varepsilon_-({\bm k}),\varepsilon_-({\bm k}))$, where $\varepsilon_{\pm}({\bm k})$ are the energies of the normal electrons in the conduction band ($\varepsilon_+$) and valence band ($\varepsilon_-$). By using the unitary operator, the Cooper pair amplitude in the spin and orbital space, $\hat{\mathcal{F}}$, is mapped to the band representation as $\hat{\mathcal{F}} \rightarrow \hat{\mathcal{F}}^{\prime}\equiv\hat{U}^{\dag}\hat{\mathcal{F}}\hat{U}^{\ast}$. Owing to the strong spin-orbit coupling in the parent material, the odd-frequency even-parity pairs in the band representation, $\hat{\mathcal{F}}^{\prime}_{E_g}$, are regarded as an admixture of the $s$-wave pair component and non-$s$-wave (e.g., the $d$-wave) pair components. Although the $s$-wave component of $E^{({\rm o})}_g$ pairs can survive in the DN region, the additional ingredients of non-$s$-wave components in the $E^{({\rm o})}_g$ pairs are not immune to nonmagnetic impurities. This results in the fragility of the anomalous proximity effect in STIs.

Cooper pair amplitudes and local DOS in the junction of the chiral state are shown in Fig.~\ref{fig:NS_F_chiral}, where the pair potential is given by $\hat{\Delta}(z) =\Delta_0(s_x+is_y)\hat{\sigma}_y\Theta(-z)$. Similarly with the nematic state, the odd-frequency even-parity ($E_{g,s_+}^{({\rm o})}$) pair penetrates into the DN region, but its amplitude is sensitive to nonmagnetic impurities. As shown in Fig.~\ref{fig:NS_F_chiral}(d), for $\gamma_{\rm N}/\Delta_0=0.2$, the local DOS in the DN region is enhanced by the proximitized $E_{g,s_+}^{({\rm o})}$ pairs. In addition, the spin-polarization of the local DOS, $\mathcal{N}_{\rm spin}\neq 0$, reflects the spin structure of $E_{g,s_+}^{({\rm o})}$ pairs. As $\gamma_{\rm N}$ increases, however, the local DOS reduces to $\mathcal{N}_{\rm F}$ and spin-polarization of the local DOS is weakened.

\begin{figure}[t!]
\includegraphics[width=85mm]{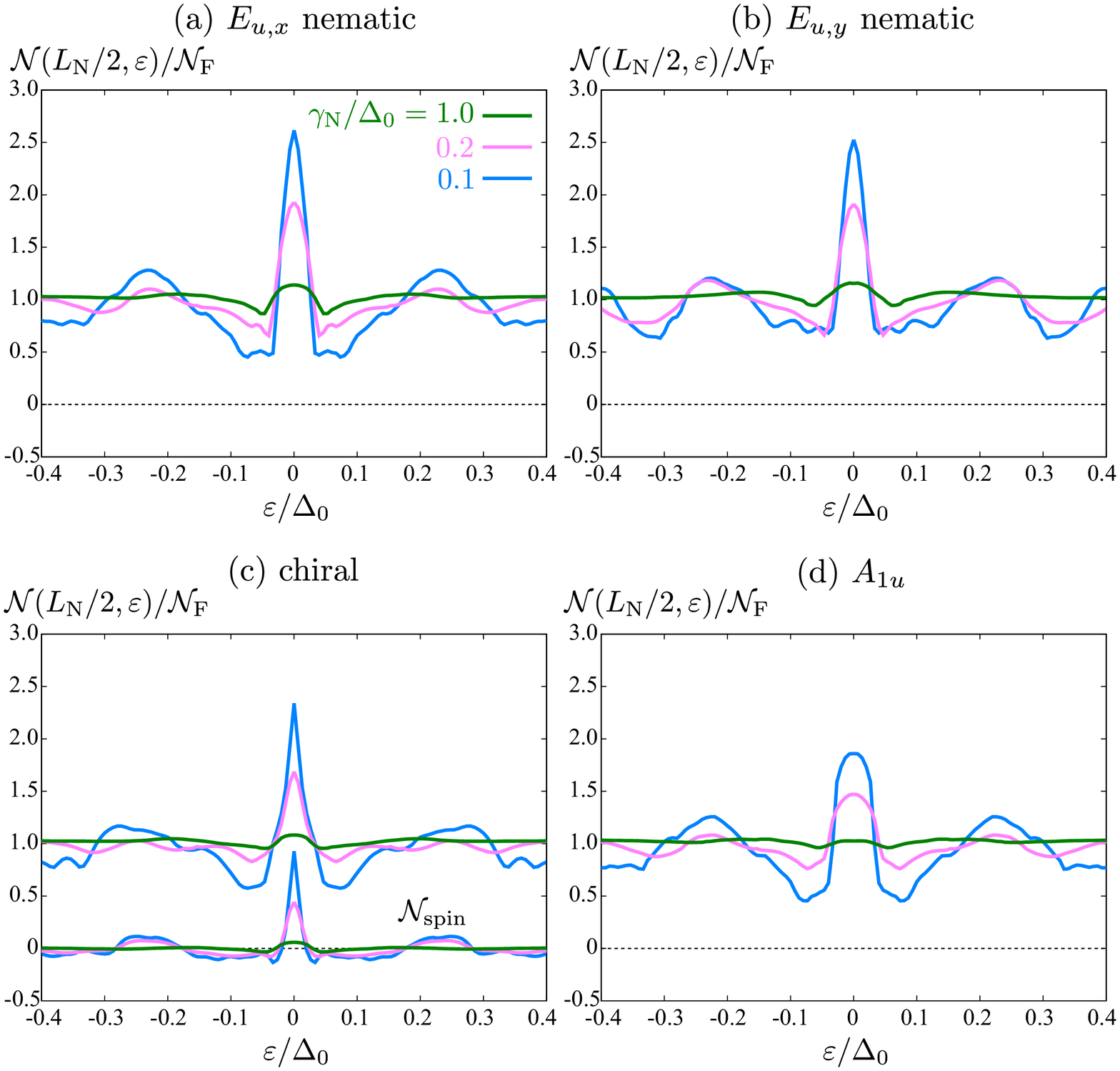}
\caption{Local DOS at $z=L_{\rm N}/2$ of the DN region: (a) the $E_{u,x}$ nematic state, (b) the $E_{u,y}$ nematic state, (c) the chiral state, and (d) the $A_{1u}$ state. In (c), we also plot the spin-resolved local DOS, $\mathcal{N}_{\rm spin}(L_{\rm N}/2,\varepsilon)$. In all data, we set $\mu/|m_0|=1.6$ and $\lambda k^2_{{\rm F},\parallel}/v= 0.1$.}
\label{fig:dos_dn}
\end{figure}

\begin{figure*}[t!]
\includegraphics[width=175mm]{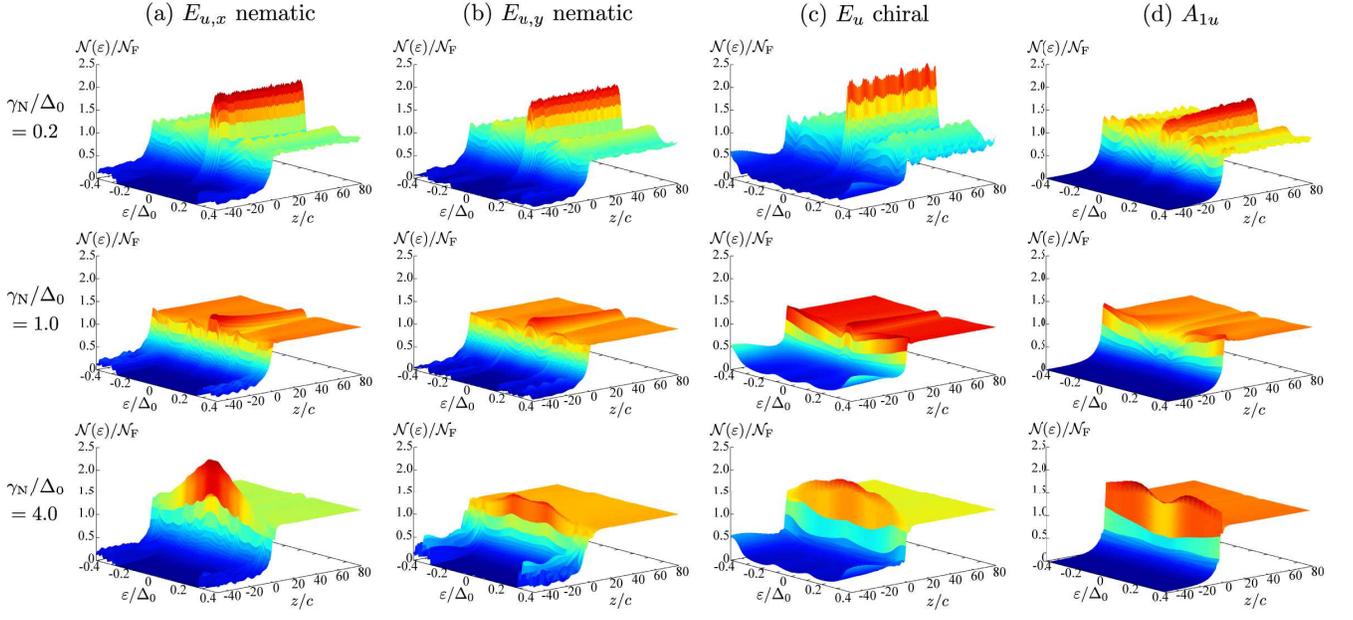}
\caption{Local DOS in STI/DN junctions for the $E_{u,x}$ nematic state (a), the $E_{u,y}$ nematic state (b), the chiral state (c), and the $A_{1u}$ state (d), where the $z\le 0$ ($z>0$) region corresponds to STI (DN). The DN is the doped topological insulators with nonmagnetic impurities, where the impurity parameter is set to be $\gamma_{\rm N}/\Delta_0=0.2$, $1.0$, and $4.0$, in the the upper, middle, and lower rows, respectively. In all data, we set $\mu/|m_0|=1.6$ and $\lambda k^2_{{\rm F},\parallel}/v= 0.1$.}
\label{fig:DN_LDOS}
\end{figure*}

In Fig.~\ref{fig:dos_dn}, we display the local DOS and spin-resolved local DOS in the DN region ($z=L_{\rm N}/2$) for the $E_{u,x}$ nematic state, the $E_{u,y}$ nematic state, the $E_u$ chiral state, and the $A_{1u}$ state. The spatial profiles of the local DOS in the vicinity of the interface are also shown in Fig.~\ref{fig:DN_LDOS}. As mentioned in previous section, the $E_{u,x}$ nematic state preserves the chiral symmetry, while the $E_{u,y}$ state spontaneously breaks the symmetry. The low-lying ABS of the latter state on the specular surface is gapped out and becomes dispersive in the surface Brillouin zone. However, we do not observe the apparent difference from the local DOS, where the both nematic states have similar profiles of the local DOS. When the impurity potential is weak, all the odd-parity states have a pronounced peak of the local DOS at $\varepsilon=0$. As $\gamma_{\rm N}$ increases, however, the peak structure reduces to $\mathcal{N}_{\rm F}$ and the characteristic structure of the local DOS disappears in the DN region.

\section{Concluding remarks}

We have studied emergent odd-frequency pairs and anomalous proximity effect in the nematic and chiral states of STIs, $M_x$Bi$_2$Se$_3$ ($M=$ Cu, Sr, Nb). We have shown how the multi-orbital degrees of freedom and strong spin-orbit coupling induce odd-frequency pairing in the bulk. The nematic and chiral states in the $E_u$ representation of the $D_{3d}$ crystalline symmetry are the prototypes of topological SCs with and without time-reversal symmetry, respectively. As a reflection of such nontrivial topology, the nematic state hosts helical Majorana fermions on the surface, while the chiral state is accompanied by spin-polarized Majorana fermions. In particular, the nematic state with the particular configuration of the nematic angle ($\hat{\bm n}\parallel \hat{\bm x}$) preserves the chiral symmetry. The chiral symmetry ensures the existence of the odd-frequency even-parity pair amplitudes at the boundary as a consequence of the spectral bulk-boundary condition, implying the topological aspect of odd-frequency even-parity pairing. In addition, the surface ABSs are orbitally polarized as a consequence of the interplay with the Dirac fermions inherent to the normal STIs. Such orbital polarization induces the parity mixing of the odd-frequency pairs on the surface, and the $E^{({\rm o})}_g$ pair amplitude associated with the SBBC necessarily leads to the large amplification of the $E^{({\rm o})}_u$ pairs. Hence, a rich variety of odd-frequency pairs emerge from the bulk and surface of the superconducting topological insulators as a reflection of the symmetry and topology of the nematic and chiral states. 

We have applied the symmetry classification to STI/normal-metal junctions, and examined the anomalous proximity effect of the odd-frequency pairs into dirty normal metals. It has been emphasized that the tunneling conductance reflects the information of normal electrons and superconducting gap symmetry; In the nematic state, the topological phase transition induced by the increase of the carrier density gives rise to the evolution of the tunneling conductance from the pronounced ZBCP to the V-shaped profile reflecting the nodal structure of the superconducting gap on the cylindrical Fermi surface. In the case of the chiral state, the non-unitary pairing induces the spin-polarization in the tunneling conductance, which discriminates the chiral state from other pairing states. In junction systems with the dirty metals, we have observed that the anomalous proximity effect in the nematic state of STIs is sensitive to nonmagnetic impurities, while the Dirac SCs without spin-orbit coupling yield anomalous proximity effect immune to nonmagnetic impurities. The odd-frequency even-parity pairs proximitized to the DN side are the admixture of $s$-wave and $d$-wave pair channels through the strong spin-orbit coupling inherent to the parent material of STIs, and the resulting anomalous proximity effect is sensitive to the strength of nonmagnetic impurities. These results imply that strong spin-orbit coupling can essentially change the properties of odd-frequency pairs. The systematic study on the effect of spin-orbit coupling remains as a future problem.

In Appendix~\ref{sec:sbbc}, we have demonstrated that the $E_{u,x}$ nematic state with the chiral symmetry maintains the SBBC, which ensures the existence of odd-frequency pair amplitudes at the boundary. The SBBC also uncovers the two different divergent behaviors of the odd-frequency pairs, which are associated with the bulk topology and symmetry. The spectral bulk-boundary correspondence may ensure the robustness of the ZBCP on the surface of the $E_{u,x}$ nematic state. Although the SBBC is recently generalized to the chiral-symmetric system with randomly distributed impurities, its relation to the anomalous proximity effect still remains an unresolved puzzle. It might be important to revisit the anomalous proximity effect in chiral-symmetric systems with/without spin-orbit coupling from the viewpoint of the SBBC.

\begin{acknowledgments} 

This work was supported by a Grant-in-Aid for Scientific Research on Innovative Areas ``Quantum Liquid Crystals (No.~JP22H04480)'' from JSPS of Japan, JSPS KAKENHI (Grant No.~JP20K03860, No.~JP20H01857, No.~JP20H00131, No.~JP21H01039, and No.~JP22H01221) and JSPS-EPSRC
Core-to-Core program ``Oxide Superspin''.

\end{acknowledgments}

\appendix

\section{Thermodynamic stability of nematic and chiral states}
\label{sec:phase}

Here we discuss the thermodynamic stability of the $E_u$ nematic and chiral states. Let us consider the interaction Hamiltonian~\cite{fuPRL10}
\beq
\mathcal{H}_{\rm int} = \int d{\bm r} 
\left[U(n^2_1({\bm r})+n^2_2({\bm r}))+2Vn_1({\bm r})n_2({\bm r})\right],
\label{eq:Hint}
\eeq 
where $n_{\sigma}\equiv \sum _{s=\uparrow,\downarrow}c^{\dag}_{s,\sigma}c _{s,\sigma}$ is the density operator in orbital $\sigma$, and $U$ and $V$ denote the intra-orbital and inter-orbital interaction constants, respectively. The $U$-$V$ model, which describes the short-range density interaction between electrons, is the simplest model to describe the competition between intra-orbital pairing ($A_{1g} $ and $A_{2u}$) and inter-orbital odd-parity pairing states ($A_{1u}$ and $E_u$) in doped Bi$_2$Se$_3$. Recently, inelastic neutron scattering measurements on Sr$_{0.1}$Bi$_2$Se$_3$ unveiled highly anisotropic acoustic phonons along the [001] direction~\cite{wan19}. Contrary to isotropic electron-phonon coupling, the observation reflects the singular electron-phonon coupling which may assist the inter-orbital pairing rather than conventional $s$-wave ($A_{1g}$) pairing. Hence we start with the $U$-$V$ model to elucidate the role of the hexagonal warping effect on the stability of the inter-orbital pairing states.

We start with the $8\times 8$ Matsubara Green's function at temperature $T$ defined in Eq.~\eqref{eq:Gorkov}. 
For spatially uniform systems, the formal solution of the Gor'kov equation is obtained as 
\beq
{\check{G}}({\bm k},i\varepsilon _n) = \begin{pmatrix}
\hat{G}({\bm k},i\varepsilon _n) & \hat{F}({\bm k},i\varepsilon _n) \\
\hat{\bar{F}}({\bm k},i\varepsilon _n) & \hat{\bar{G}}({\bm k},i\varepsilon _n)
\end{pmatrix}
= \left[
i\varepsilon _n - {\check{\mathcal{H}}}({\bm k})
\right]^{-1} .
\label{eq:Gk}
\eeq
The Green's function is determined by self-consistently solving Eq.~\eqref{eq:Gk} with the gap equation,
\beq
\hat{\Delta}({\bm r})=-i\mathcal{V} T\sum_n \hat{F}({\bm r},{\bm r},i\varepsilon_n) s_y.
\label{eq:gapeq}
\eeq
Here we consider the interaction Hamiltonian in Eq.~\eqref{eq:Hint}, which is composed of a contact and attractive intera-orbital and inter-orbital interactions. Hence, the coupling constant in Eq.~\eqref{eq:gapeq} is set to be $\mathcal{V}=U$ for intra-orbital channels and $\mathcal{V}=V$ for inter-orbital channel. 

Let $\hat{G}_{\rm N}({\bm k},i\varepsilon _n) = [i\varepsilon _n - \hat{h}({\bm k})]^{-1}$ be the Green's function in the normal state, where $\hat{h}$ is introduced in Eq.~\eqref{eq:hti}. The normal state Hamiltonian are diagonalized as 
$\hat{h}({\bm k})\ket{u_{\pm}({\bm k})}=\varepsilon_{\pm}({\bm k})\ket{u_{\pm}({\bm k})}$, where the dispersions of the conduction band ($E_+$) and valence band ($E_-$) are 
\begin{align}
\varepsilon_{\pm}({\bm k})
=&c({\bm k})-\mu \nn \\
& \pm \sqrt{m^2({\bm k})+(v_zf_z)^2+v^2(f^2_x+f^2_y)+\lambda^2f_{3\lambda}^2}.
\label{eq:epm}
\end{align}
The Green's function in the normal state, $\hat{G}_{\rm N}$, is given by
\beq
\hat{G}_{\rm N}({\bm k},i\varepsilon_n) = \frac{\mathcal{I}_{+}({\bm k})}{i\varepsilon _n - \varepsilon_+({\bm k})}
+  \frac{\mathcal{I}_{-}({\bm k})}{i\varepsilon _n - \varepsilon_-({\bm k})}.
\eeq
where the projection operator onto the conduction (valence) band, $\mathcal{I}_+({\bm k})$ ($\mathcal{I}_-({\bm k})$) is defined as 
\beq
\mathcal{I}_{\pm} \equiv \ket{u_{\pm}({\bm k})}  
\bra{u_{\pm}({\bm k})}
+ \mathcal{P}\mathcal{T}\ket{u_{\pm}({\bm k})} 
\bra{ u_{\pm}({\bm k})}(\mathcal{PT})^{-1}.
\eeq
The anomalous Green's function is then expanded in terms of $\hat{G}_{\rm N}$ as  
\beq
\hat{F}({\bm k},i\varepsilon _n)=\hat{G}_{\rm N}({\bm k},i\varepsilon _n)\hat{\Delta} ({\bm k})\hat{\bar{G}}({\bm k},i\varepsilon _n).
\eeq

\begin{figure}[t]
\includegraphics[width=85mm]{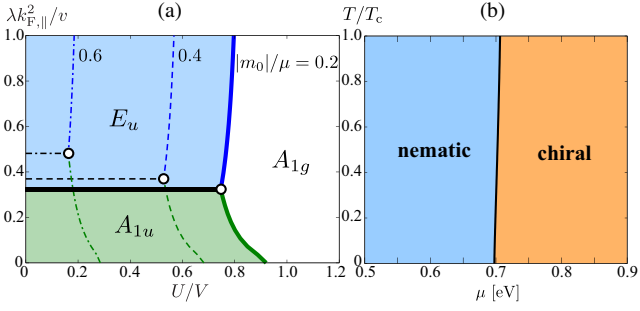}
\caption{(a) Phase diagram in the plane of the pair interaction $U/V$ and the strength of the hexagonal warping term, where we set $|m_0|/\mu=0.2$ (solid lines), $0.4$ (dashed lines), and $0.6$ (dotted-dashed lines). (b) Phase diagram within the $E_u$ representation, where we set $\lambda = 0$.}
\label{fig:phase}
\end{figure}

At the superconducting transition temperature, $T=T_{\rm c}$, the Green's function can be expanded as $\hat{G}({\bm k},i\varepsilon_n)=\hat{G}_{\rm N}({\bm k},i\varepsilon_n)+\mathcal{O}(\Delta/T_{\rm c})$. To obtain the phase diagram in Fig.~\ref{fig:phase}, therefore, we linearize the gap equation by replacing $\hat{\bar{G}}({\bm k},i\varepsilon_n)$ with $-s_y\hat{G}_{\rm N}({\bm k},-i\varepsilon _n)s_y$. Let $\hat{d}^{\Gamma}_j$ be the basis function of the irreducible representation of $D_{3d}$, i.e., $(\hat{d}^{A_{1g}}_1, \hat{d}^{A_{1g}}_2) =(1,\sigma _x)$ for the $A_{1g}$ state, $\hat{d}^{A_{1u}} =\sigma _y s_z$ for the $A_{1u}$ state, $\hat{d}^{A_{2u}} = \sigma _z$ for the $A_{2u}$ state, and $(\hat{d}^{E_u} _1, \hat{d}^{E_u} _2)=(\sigma _ys_x,\sigma _ys_y)$ for the $E_u$ state. We now expand the pair potential in terms of $\hat{d}^{\Gamma}_i$ as $\Delta = \sum_j \eta_j \hat{d} _j^{\Gamma}$, where $\eta _j \in \mathbb{R}$ without loss of generality. The linearized gap equation for the $\Gamma$ representation is given as
\begin{align}
\eta _i =\sum _j \chi^{\Gamma}_{ij}(T)\eta _j,
\end{align}
where the susceptibilities in each channel are obtained by substituting $\hat{F}=\hat{G}_{\rm N}\hat{\Delta} \hat{\bar{G}}$ into the gap equation \eqref{eq:gapeq} as 
\begin{align}
\chi^{\Gamma}_{ij}(T) = - \frac{1}{4} T \sum _n \sum_{\bm k}{\rm tr} \left[
\hat{d}^{\Gamma}_i \mathcal{V} \hat{G}_{\rm N}({\bm k},-i\varepsilon _n)
\hat{d}^{\Gamma}_j \hat{G}_{\rm N}({\bm k},i\varepsilon _n)  \right].
\label{eq:homo}
\end{align}
We have introduced the coupling constants, $\mathcal{V} = U$ for intraorbital interaction and $\mathcal{V} = V$ for interorbital interaction. The superconducting susceptibility tensor, $\chi _{ij}$, is reduced to
\beq
\chi^{\Gamma}_{ij} = - \frac{1}{4}
\int \frac{d^3{\bm k}}{(2\pi)^3}
\frac{{\rm tr}[\hat{d}^{\Gamma} _i \mathcal{M}_j({\bm k})]}{2\varepsilon_+({\bm k})}
\tanh\left( \frac{\varepsilon_+({\bm k})}{2T}\right),
\eeq
where $[\hat{\mathcal{M}}_j({\bm k})]_{\alpha\beta} \equiv \mathcal{V}_{\alpha\beta} [\mathcal{I}_+({\bm k})\hat{d}^{\Gamma} _j \mathcal{I}_+({\bm k})]_{\alpha\beta}$. The nontrivial solution of the linear homogeneous equation \eqref{eq:homo} is determined by 
\beq
{\rm det} \left( \underline{1} - \underline{\chi}(T_{\rm c})\right) = 0.
\eeq
This is the linearized equation for determining the superconducting transition temperature $T_{\rm c}$ in each pairing channel.

It is seen from Fig.~\ref{fig:phase}(a) that the $A_{1g}$ phase enlarges in small $\lambda$ region. The phase boundary between the $A_{1g}$ and $A_{1u}$ states is indeed given by
\beq
\frac{U}{V} = 1 - \frac{2m^2_0}{\mu^2} - \lambda^2 \left|\delta (T_{\rm c})\right|,
\eeq
where 
\beq
\delta (T_{\rm c}) = \frac{1}{\chi _0(T_{\rm c})} \int \frac{d^3{\bm k}}{(2\pi)^3}\frac{(k^3_++k^3_-)^2}{E_{0}+\mu}\frac{\tanh(E_0/2T_{\rm c})}{2E_0},
\eeq
with $E_0$ being the dispersion of the conduction band for $\lambda = 0$ and 
$\chi _0(T_{\rm c}) = - \frac{1}{2}\int \frac{d^3{\bm k}}{(2\pi)^3}\tanh(E_0/2T_{\rm c})/2E_0<0$. The leading order correction of the hexagonal warping effect, $\delta (T_{\rm c})$, turns out to be negative definite $\delta (T_{\rm c})<0$.
The deviation, $\delta (T_{\rm c})$, originates from the leading order correction of the hexagonal warping effect. The warping correction narrows the stable region of the $A_{1u}$ state against the $A_{1g}$ state. 
The leading order correction spreads the phase boundary between the $E_u$ state and the $A_{1g}$ state as 
\beq
\frac{U}{V} = \frac{2}{3} - \frac{5}{3}\frac{m^2_0}{\mu^2} + \frac{1}{2}\lambda^2 \left|\delta (T_{\rm c})\right|.
\eeq
Hence, the warping correction stabilizes the $E_{u}$ state relative to the other gapped states, such as the $A_{1g}$ and $A_{1u}$ states.

In Fig.~\ref{fig:phase}(a), we show the phase diagram in the plane of the pair interaction $U/V$ and the strength of the hexagonal warping term. We compute the critical temperature $T^{(\Gamma)}_{\rm c}$ in each irreducible representation by solving the linearized gap equation. We take $|m_0|/\mu = 0.2$ and $v=v_z$, corresponding to a spheroidal Fermi surface. The value of the interorbital interaction constant $V$ is set to be $T^{E_u}_{\rm c}=0.001T_{\rm F}$ at $\lambda = 0$. For $\lambda =0$, the $A_{1u}$ odd parity pairing can be stabilized in the region of $U/V < 1-2m^2_0/\mu^2$~\cite{fuPRL10}.  

As $\lambda$ in Eq.~\eqref{eq:hti} increases, however, the $E_u$ nematic state is fully gapped, and can become energetically competitive to the other gapped $A_{1g}$ and $A_{1u}$ states. The stability is directly attributed to the spin structure of electrons on the Fermi surface. When the warping term is absent, the electron states have helical spin textures on the basal plane, which favors the $A_{1u}$ state for $|V|\gtrsim |U|$. In contrast, the hexagonal warping term aligns the spin texture to the out-of-plane ($z$) direction, and as $\lambda$ increases, the threefold rotational spin configuration favors the $E_u$ state $c _{\uparrow,1}c_{\uparrow,2}\pm c _{\downarrow,1}c _{\downarrow,2}$. In the regime of the moderate hexagonal warping, therefore, the $E_u$ nematic states are competitive to the $A_{1u}$ state. In Fig.~\ref{fig:phase}(a), the $A_{1u}$ state indeed appears in the regime of $\lambda k^2_{{\rm  F},\parallel}/v \lesssim 1$, while the $E_{u}$ nematic state becomes stable as the hexagonal warping term increases. We note that the nematic state with $\hat{\bm n}\parallel\hat{\bm y}$ breaks the mirror reflection symmetry and its low-lying excitation becomes gapfull. 

The Fermi surface evolution from a closed spheroidal to open cylindrical shape causes the phase transition from the nematic to chiral state. To clarify this, we introduce the effective pairing interaction, 
\beq
V_{ab;cd}({\bm k},{\bm k}^{\prime}) = - \sum _{j=1}^{n_{\Gamma}} V^{\Gamma}_j \left(i\hat{d}^{(\Gamma)}_j({\bm k}) s_y\right)_{ab} \left(i\hat{d}^{(\Gamma)}_j({\bm k}^{\prime})s_y\right)^{\ast}_{cd} ,
\eeq
where $V^{\Gamma}_j>0$ is the effective coupling constant in the irreducible representations $\Gamma=\{ A_{1g}, A_{2g}, A_{2u}, E_u \}$ of the point group $D_{3d}$ with the dimension $n_{\Gamma}$ and basis functions $\{ \hat{d}^{(\Gamma)}_j({\bm k}) \}_{j=1,\cdots, n_{\Gamma}}$. To focus on the nematic-to-chiral phase transition, we assume the $E_{u}$ channel as the dominant pairing interaction and set $V^{A_{1g}}=V^{A_{1u}}=V^{A_{2u}}=0$. Employing the quasiclassical approximation for electrons in the conduction band, we compute the thermodynamic potential and determine the thermodynamically stable phase under $T$ and $\mu$.  The first-order nematic-to-chiral phase transition occurs around $\mu = 0.7~{\rm eV}=2.5|m_0|$. In such high carrier density with a two-dimensional cylindrical Fermi surface, the gap function of the chiral state in the band representation [Eq.~\eqref{eq:dchiral}] reduces to that of the two-dimensional chiral $p$-wave state, ${\bm d}({\bm k})\approx[0,0,v(k_x+ik_y)]/|m_0|$. The fully gapped chiral state is energetically favored in high carrier density regime, compared with the nematic state, ${\bm d}({\bm k})\approx (0,0,vk_x)/|m_0|$, having the line nodes along the $k_z$ direction. We note that the phase boundary in Fig.~\ref{fig:phase}(b) is less sensitive to the strength of the hexagonal warping term, as the stability of the chiral state is attributed to the gain of the condensation energy due to the Fermi surface evolution.

Lastly, we mention the stability of the odd-parity $E_u$ states. It has been pointed out that the critical temperature of the odd-parity $A_{1u}$ and $E_u$ states is sensitive to nonmagnetic impurities~\cite{nagaiPRB14,sato20}. This is contrast to the case of the even-parity $A_{1g}$ state which is basically robust against nonmagnetic impurities. Recently, twofold rotational symmetry was reported above $T_{\rm c}$ in Sr$_x$Bi$_2$Se$_3$~\cite{kunNJP18,sunPRL19}. This indicates that the $E_u$ nematic state can be stabilized by the explicit breaking of the threefold symmetry in the basal plane of the $D_{3d}$ crystal~\cite{kunNJP18,sunPRL19,kun19} or externally applied uniaxial stress~\cite{kos19}. As the non-unitary chiral state in the $E_u$ representation is accompanied by magnetization at zero fields,  magnetic impurities can also help to stabilize the chiral state~\cite{yuanPRB17}.

\section{Spectral bulk-boundary correspondence in nematic states}
\label{sec:sbbc}

In chiral symmetric systems, the Hamiltonian density, $\check{\mathcal{H}}({\bm k})$, is the anticommutable with the chiral operator $\check{\Gamma}$. According to the bulk-boundary correspondence reflects the intrinsic relation between the nontrivial topology of the bulk and boundary states. The winding number defined with chiral operator predicts the number of the zero energy states at boundaries~\cite{satoPRB11}. In addition, the SBBC, which is a generalization of bulk-boundary correspondence in one-dimensional chiral symmetric systems into complex frequencies, reflects the intrinsic relation between the bulk quantity evaluated over the entire frequency range and and the odd-frequency Cooper pairs accumulated at the boundary~\cite{tamuraPRB19,tanaka21}. 
The SBBC was proved by using an analogy to the concept of electric polarization~\cite{daiPRB19}, and recently generalized to systems with impurities and dynamical self-energies~\cite{tamura21}. Here we numerically demonstrate that the $E_{u,x}$ nematic state in STIs without nonmagnetic impurities, which holds the chiral symmetry in the $k_y$-$k_z$ plane, satisfies the SBBC and the characteristic critical behaviors of the odd-frequency pairs.

Let us recall the chiral operator in Eq.~\eqref{eq:chiral}, which is defined as a combination of the mirror reflection symmetry, the time-reversal symmetry, and the particle-hole symmetry, and anticommutable with the BdG Hamiltonian,
\beq
\{ \check{\Gamma} _1, \check{\mathcal{H}} (0,k_y,k_z)\} = 0.
\eeq
In the chiral symmetric $k_y$-$k_z$ plane, i.e., $k_x=0$, the SBBC is then explicitly written as 
\beq
F^{\rm SBBC}(k_y,\omega) = \frac{W(k_y,\omega)}{\omega}.
\label{eq:sbbc}
\eeq
The quantity in the left hand side is defined as
\beq
F^{\rm SBBC}(k_y,\omega) \equiv {\rm Tr}_j \left[
\check{\Gamma}_1 \check{G}(k_y,\omega)
\right],
\label{eq:Fsbbc}
\eeq
where $\check{G}_{ij}(k_y,\omega)\equiv[\omega-\check{H}(z_i,z_j,k_x=0,k_y)]^{-1}$ is the Green's function with the open boundary condition along the $z$ axis and the trace is taken over a semi-infinite space: ${\rm Tr}_j \cdots \equiv {\rm tr}\sum _j \langle j| \cdots |j\rangle$. The quantity $F^{\rm SBBC}$ is an odd function on $\omega$
\beq
F^{\rm SBBC}(k_y,\omega) = -F^{\rm SBBC}(k_y,-\omega),
\eeq
which is responsible for ZBCPs in tunneling conductance and anomalous proximity effect as shown in Sec.~\ref{sec:proximity}.
In superconducting states, where the chiral operator is constructed from the particle-hole operator, therefore, the left hand side of Eq.~\eqref{eq:sbbc} represents the odd-frequency Cooper pairs accumulated at the boundary. 
The right hand side of Eq.~\eqref{eq:sbbc} is evaluated from the bulk Green's function, $\check{G}({\bm k},\omega) = [\omega -\check{\mathcal{H}}({\bm k})]^{-1}$, as 
\begin{align}
W(k_y,\omega) = -\int^{+\pi/c}_{-\pi/c}\frac{dk_z}{4\pi i}{\rm tr}
\left[\check{\Gamma} \check{G}({\bm k},\omega)\partial _{k_z} \check{G}^{-1}({\bm k},\omega)\right]_{k_x=0}.
\end{align}
This is the generalization of the one-dimensional winding number to the complex frequency plane.

We note that $W(\omega)$ can be well defined only when a finite energy gap is opened in the momentum plane invariant under the chiral symmetry, i.e., the $k_y$-$k_z$ plane. Then, the bulk quantity $W(\omega)$ can be expanded in terms of the small value of $\omega$ as $W(\omega) = \sum_{l\ge 0}W_{l}\omega^l$, as far as $|\omega|$ is sufficiently smaller than the energy gap of the quasiparticle spectrum in the chiral-symmetric momentum space. From the Laurent expansion on the complex $\omega$ plane and Eq.~\eqref{eq:sbbc}, the odd frequency pair amplitude accumulated at the boundary is expressed in the low-frequency limit as 
\beq
F^{\rm SBBC}(k_y,\omega)=\frac{w_{\rm 1d}(k_y)}{\omega}+\chi(k_y)\omega+O(\omega^3). 
\label{eq:singular}
\eeq
The zeroth-order coefficient $W_{l=0}$ corresponds to the topological invariant associated with the chiral symmetry, 
\beq
W_{l=0}=\lim _{\omega\rightarrow 0}W(k_y,\omega) = w_{\rm 1d}(k_y).
\eeq
Thus $W(k_y,\omega)$ reduces to the one-dimensional winding number in Eq.~\eqref{eq:w1d} at $\omega\rightarrow 0$, and Eq.~\eqref{eq:sbbc} connects the nontrivial topological invariant $w_{\rm 1d}(k_y)\in \mathbb{Z}$ to the odd-frequency pair through the singular functional form $\lim_{\omega\rightarrow 0}F^{\rm SBBC}(k_y,\omega)=w_{\rm 1d}(k_y)/\omega$ at $\omega\rightarrow 0$. It has also been shown that the coefficient of the next-leading-order term, 
\beq
\chi(k_y)\equiv W_{l=1}= \frac{1}{2}\frac{\partial^2W(k_{y},\omega)}{\partial \omega^2}\bigg|_{\omega\rightarrow 0}, 
\eeq
yields a power law divergence at the topological quantum phase transition. Hence, Eq.~\eqref{eq:singular} describes two different types singularities of odd-frequency pair amplitudes protected by the chiral symmetry, the $1/\omega$ divergence in the topological phase with $w_{\rm 1d}(k_y)\neq 0$ and a power law divergence of $\chi (k_y)$ at the topological quantum phase transition.

Here we numerically demonstrate the SBBC in Eq.~\eqref{eq:sbbc} and the critical behavior of $\chi(k_y)$ in the $E_{u,x}$ nematic state of STIs. We note that in the $E_{u,x}$ nematic state, $|w_{\rm 1d}(k_y)|=2$ for $|k_{y}|<k_{{\rm F},y}$ protects the existence of the flat-band zero-energy states along $k_y$ and the end points $k_y = \pm k_{{\rm F},y}$ correspond to the nodal points of the bulk superconducting gap at which $W$ is not well-defined. We consider the system same as that in Sec.~\ref{sec:surface}, which is described by the BdG Hamiltonian in Eq.~\eqref{eq:Hbdg} with open boundary along $z$ direction and periodic boundary conditions in the $x$-$y$ plane.

\begin{figure}[t!]
\includegraphics[width=85mm]{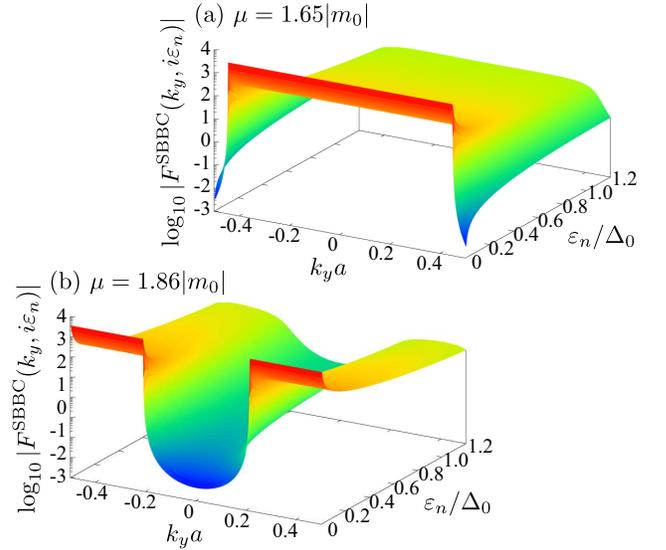}
\caption{Odd-frequency pair amplitudes associated with the SBBC, $F^{\rm SBBC}(k_y,i\varepsilon_n)$, in the $E_{u,x}$ nematic state with a closed Fermi surface $\mu = 1.65|m_0|$ (a) and an opened Fermi surface $\mu = 1.86|m_0|$ (b). $F^{\rm SBBC}$ is defined in Eq.~\eqref{eq:sbbc}. In all data, we set $\lambda k^2_{{\rm F},\parallel}/v =0.1$.}
\label{fig:sbbc1}
\end{figure}

\begin{figure}[t!]
\includegraphics[width=85mm]{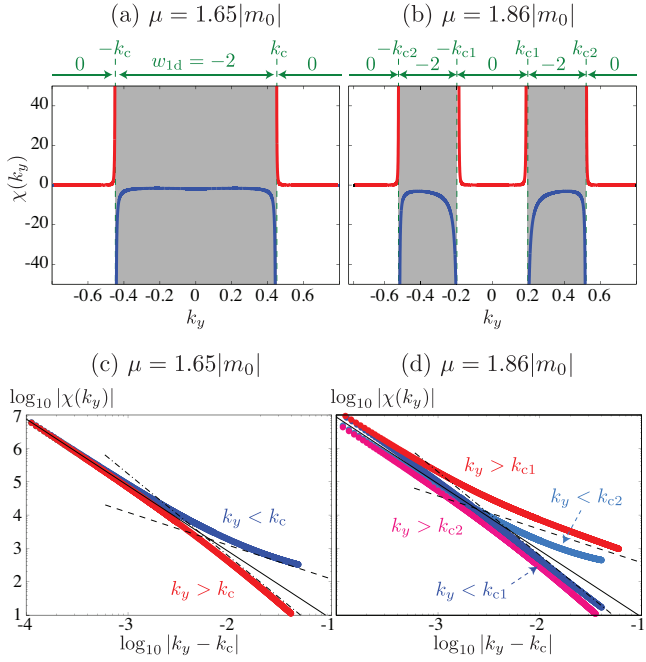}
\caption{(a,b) The bulk quantity $\chi(k_y)$ and the winding number $w_{\rm 1d}(k_y)$ in the $E_{u,x}$ nematic state as a function of $k_y$: (a) $\mu = 1.65|m_0|$ and (b) $1.86|m_0|$. The zero-energy flat bands appear in the momentum segments within $|k_y |<k_{\rm c}\approx 0.45a^{-1}$ in (a) and $k_{\rm c1}\approx 0.2 a^{-1}<k_y<k_{\rm c2}\approx 0.5a^{-1}$ and $-k_{\rm c2}<k_y<-k_{\rm c1}$ in (b). (c,d) The critical behavior of $|\chi(k_y)|$ close to the topological phase transition points $k_{\rm c}$ (c) and $k_{\rm c1}$ and $k_{\rm c2}$ (d). In (c, d), the solid, dashed and dotted-dashed lines correspond to $|k-k_{\rm c1,c2}|^{\alpha}$ with $\alpha = -2$, $-1$, and $-5/2$, respectively. 
In all data, we fix $\lambda k^2_{{\rm F},\parallel}/v = 0.1$.}
\label{fig:sbbc2}
\end{figure}

Figure~\ref{fig:sbbc1} shows the odd-frequency pair amplitudes associated with the SBBC, $F^{\rm SBBC}(k_y,i\varepsilon_n)$, for a closed Fermi surface ($\mu = 1.65|m_0|$) and an opened Fermi surface ($\mu = 1.86|m_0|$), where we take pure imaginary frequencies $\omega = i\varepsilon_n$ ($\varepsilon_n \in \mathbb{R}$). The pair amplitude, $F^{\rm SBBC}(k_y,i\varepsilon_n)$, is obtained from Eq.~\eqref{eq:Fsbbc} with the Green's function computed in the finite system. Note that the pair amplitudes associated with the SBBC is equivalent to the odd-frequency even-parity ($E_g$) pair amplitudes, 
\beq
F^{\rm SBBC}(k_y,\omega)  = F^{\rm OTEE}_{E_g}(k_y,\omega).
\eeq
In the case of the closed Fermi surface ($\mu = 1.65|m_0|$), the winding number is nontrivial $w_{\rm 1d}(k_y)=-2$ and the zero-energy flat bands appear in the momentum segment $|k_y|<k_{\rm c}\approx 0.45a^{-1}$ [see also Fig.~\ref{fig:sbbc2}(a)]. As shown in Fig.~\ref{fig:sbbc2}(b), the topologically nontrivial momentum region splits to two parts, $k_{\rm c1}\approx 0.2 a^{-1}<k_y<k_{\rm c2}\approx 0.5a^{-1}$ and $-k_{\rm c2}<k_y<-k_{\rm c1}$, in the opened Fermi surface.
Figures~\ref{fig:sbbc1}(a) and \ref{fig:sbbc1}(b) show the divergent behavior of $F^{\rm SBBC}(k_y,i\varepsilon_n)$ at $\varepsilon_n\rightarrow 0$ in topologically nontrivial momentum segments. We calculate the bulk quantity $W(k_y,\omega)$ and confirm that Eq.~\eqref{eq:sbbc} holds within numerical accuracy $|i\varepsilon_nF^{\rm SBBC}(k_y,i\varepsilon_n)-W(k_y,i\varepsilon_n)|\lesssim 10^{-8}$ for any values of $k_y$ except the vicinity of the nodal points.

We also plot in Figs.~\ref{fig:sbbc2}(a) and \ref{fig:sbbc2}(b) the bulk quantity $\chi(k_y)$ and the winding number $w_{\rm 1d}(k_y)$ in the $E_{u,x}$ nematic state as a function of $k_y$. The bulk quantity $\chi(k_y)$ exhibits a power law divergence at which the winding number $w_{\rm 1d}$ changes. As shown in Fig.~\ref{fig:sbbc2}(c), the critical behavior at $k_{y}\rightarrow k_{\rm c}$ is well describable with $\chi(k_y) \sim |k_y-k_{\rm c}|^{\alpha}$ with $\alpha=-2$ (solid line), which is consistent with the Ising universality~\cite{chaikin}. As $k_y$ moves away from $k_{\rm c}$, however, the exponent crossovers into $\alpha = -1$ for $k_{y}<k_{\rm c}$ (dashed line) and $\alpha = -5/2$ for $k_y > k_{\rm c}$ (dotted-dashed line).  The similar behaviors of $\chi(k_y)$ are observed in $\mu = 1.86|m_0|$ [Fig.~\ref{fig:sbbc2}(d)], where the zero energy flat bands appear in two different segments on $k_y$. The asymmetry of the critical exponent $\alpha$ is consistent with that of the Kitaev chain and Rashba nanowire models~\cite{tamuraPRB19}. The asymmetric exponents $\alpha = -1$ and $-5/2$ indicate that the topological critical phenomena associated with the chiral symmetry can not be categorized to ordinary Ising universality class~\cite{tamuraPRB19}.

\section{Anomalous proximity effect in Dirac SCs}
\label{sec:dirac}

The $E_{u,x}$ nematic state in STIs preserves the time-reversal symmetry and has a pair of nodal points without Berry curvatures. Similarly with the nematic SCs, Dirac SCs are the time-reversal-invariant spin-triplet $p$-wave superconducting state with a pairwise nodal points and accompanied by zero-energy surface flat band protected by the chiral symmetry. As a comparison with the nematic state in STIs, therefore, we discuss the anomalous proximity effect in Dirac SCs without spin-orbit coupling. For simplicity, we consider a tight-binding model for spin-$1/2$ electrons on the cubic lattice,
\beq
\hat{h}({\bm k}) = - 2t\sum_{i=x,y,z} \cos(k_i a)-\mu,
\eeq
where $a$ is the lattice constant. In the numerical calculation, we fix $\mu = -5t$, where the closed Fermi surface encloses the $\Gamma$ point. 
The $2\times 2$ pair potential in the spin space is represented by the ${\bm d}$-vector, which is the order parameter of spin-triplet odd-parity superconducting state, 
\beq
\hat{\Delta}({\bm k}) = i \Delta_0[{\bm s}\cdot{\bm d}({\bm k}) ]s_y.
\eeq
As a prototype of Dirac SCs, we here consider the following ${\bm d}$-vector, 
\beq
{\bm d}({\bm k}) = (\sin(k_za),\sin(k_xa),0).
\label{eq:dirac_d}
\eeq
This has a pair of nodal points at $k_y = \pm k_{{\rm F},y}$, where $k_{{\rm F},y}$ is the Fermi momentum on the $k_y$ axis. The zero-energy flat band exists along the $k_y$ axis in the $x$-$y$ surface Brillouin zone, which is similar nodal and spectral structures with those of the $E_{u,x}$ nematic state in STIs. We note that as the $x$ component of the ${\bm d}$-vector, $d_x\propto \sin(k_za)$, breaks the translational symmetry in the junction, the component of the anomalous Green's function, 
\beq
F_z(z,z,k_x,k_y,i\varepsilon_n)\equiv {\rm tr}[s_z\check{\tau}_y\check{G}(z,z,k_x,k_y,i\varepsilon_n)], 
\label{eq:Fz}
\eeq
is only the nonvanishing amplitude of the odd-frequency even-parity pair at the interface. As mentioned below, $F_z$ contains the pair amplitude protected by the chiral symmetry which is responsible for the SBBC.

\begin{figure}[t!]
\includegraphics[width=85mm]{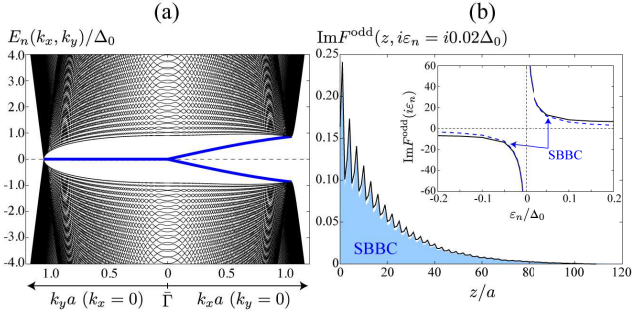}
\caption{(a) Quasiparticle spectrum on the $z$-surface of the Dirac SC with Eq.~\eqref{eq:dirac_d}, where the point nodes exist at $k_y = \pm k_{{\rm F},y}$. (b) Spatial profiles of odd-frequency pair amplitude ${F}^{\rm odd}(z,i\varepsilon_n)$ (solid line) and pair amplitude protected by the chiral symmetry $\mathcal{F}^{\rm SBBC}(z,i\varepsilon_n)$ (shaded area) at $\varepsilon_n/\Delta_0 =0.02$. The inset shows ${F}^{\rm odd}(i\varepsilon_n)\equiv \sum_j{F}^{\rm odd}(z_j,i\varepsilon_n)$ (solid line) and $\mathcal{F}^{\rm SBBC}(i\varepsilon_n)\equiv \sum_j\mathcal{F}^{\rm SBBC}(z_j,i\varepsilon_n)$ (dashed line) accumulated at the surface as a function of $\varepsilon_n$.
}
\label{fig:helical_disp}
\end{figure}

We first show the surface ABSs in the Dirac SC with the ${\bm d}$-vector given by Eq.~\eqref{eq:dirac_d}, where we consider a semi-infinite system with a specular surface at $z=0$ and periodic boundary conditions in the $x$-$y$ plane. The low-lying quasiparticle spectrum is displayed in Fig.~\ref{fig:helical_disp}(a). The zero-energy flat band appears along the $k_y$-axis and the dispersion of the low-lying surface ABSs is given by $E(k_x,k_y) = \pm \Delta_0k_x/k_{\rm F}$. In the case of the ${\bm d}$-vector with Eq.~\eqref{eq:dirac_d}, the chiral operator $\check{\Gamma}$ is defined as a combination of the time-reversal symmetry, mirror reflection symmetry on the $y$-$z$ plane ($\check{\mathcal{M}}_{yz}=i{s}_x\check{\tau}_z$), and the particle-hole symmetry, $\check{\Gamma} = {s}_z\check{\tau}_y$, which obeys $\{\check{\Gamma},\check{\mathcal{H}}(0,k_y,k_z)\}=0$. The one-dimensional winding number $w_{\rm 1d}(k_y)$ is evaluated from Eq.~\eqref{eq:w1d} as $w_{\rm 1d}=-2$ for $|k_y|< k_{{\rm F},y}$ and $w_{\rm 1d}=0$ for $|k_y|> k_{{\rm F},y}$, which ensures the existence of the zero-energy flat band along $k_y$. The chiral operator is also responsible for the SBBC in Eq.~\eqref{eq:sbbc}, which is accompanied by the chiral-symmetry-protected odd-frequency pairs, 
\begin{align}
F^{\rm SBBC}(k_y,\omega) =& \sum_{j}{\rm tr}[\check{\Gamma}\check{G}(z_j,z_j,k_x=0,k_y,\omega)] \nn \\
=& \sum_j F_z(z_j,z_j,k_x=0,k_y,\omega).
\end{align}
$F_z$ is the odd-frequency pair amplitude defined in Eq.~\eqref{eq:Fz}. We numerically confirmed that the Dirac SC in a semi-infinite system satisfies the SBBC within numerical accuracy and $F^{\rm SBBC}(k_y,\omega)$ has two different divergent behaviors at $\omega\rightarrow 0$ and $k_{y}\rightarrow \pm k_{{\rm F},y}$. 
In Fig.~\ref{fig:helical_disp}(b), we plot the spatial profiles of the odd-frequency even-parity pair amplitude at $\varepsilon_n/\Delta_0 =0.02$,
\begin{align}
{F}^{\rm odd}(z,i\varepsilon_n)\equiv \frac{1}{2}
\sum_{k_x,k_y}& \left[{F}_z(z,z,k_x,k_y,i\varepsilon_n)\right. \nn \\
& \left. -F_z(z,z,k_x,k_y,-i\varepsilon_n)\right].
\end{align}
As mentioned above, this contains the chiral-symmetry-protected pair amplitude. Thus we also plot in Fig.~\ref{fig:helical_disp}(b) the spatial profile of the chiral-symmetry-protected pair amplitude (shaded area),
\beq
\mathcal{F}^{\rm SBBC}(z,i\varepsilon_n)\equiv \sum_{k_y}F_z(z,z,k_x=0,k_y,\omega).
\eeq 
Since the pair amplitude is a purely imaginary function, we do not show the real part of $F$ in Fig.~\ref{fig:helical_disp}(b). At $\varepsilon_n \rightarrow 0$, the odd-frequency pair amplitude localized at the boundary is solely composed of $\mathcal{F}^{\rm SBBC}(z,i\varepsilon_n)$ and protected by the chiral symmetry. The inset of Fig.~\ref{fig:helical_disp}(b) plots the odd-frequency pair amplitude accumulated at the boundary, $F^{\rm odd}(i\varepsilon_n)=\sum_jF^{\rm odd}(z_j,i\varepsilon_n)$, as a function of $\varepsilon_n$ (solid lines). It diverges in the limit $\varepsilon_n\rightarrow 0$, and the singularity is attributed to $\mathcal{F}^{\rm SBBC}(i\varepsilon_n)\equiv \sum_j\mathcal{F}^{\rm SBBC}(z_j,i\varepsilon_n)$ (dashed lines) obeying the SBBC, $\mathcal{F}^{\rm SBBC}(i\varepsilon_n\rightarrow 0)= \sum _{|k_y|<k_{{\rm F},y}}w_{\rm 1d}(k_y)/i\varepsilon_n \propto 1/i\varepsilon_n$.

\begin{figure}[t!]
\includegraphics[width=87mm]{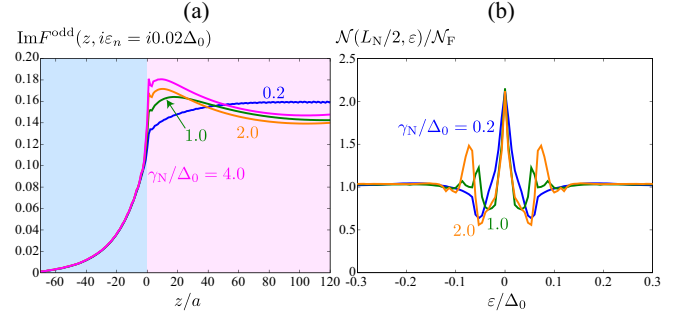}
\caption{(a) Spatial profiles of odd-frequency even-parity pair amplitudes $F^{\rm odd}(z,i\varepsilon_n)$ in the Dirac SC/DN junction for $\gamma_{\rm N}/\Delta_0=0.2$, 1.0, 2.0, and 4.0, where $z<0$ ($z>0$) corresponds to the SC (DN) region. We fix $\varepsilon_n = 0.02\Delta_0$. (b) Local DOS in the DN region ($z=L_{\rm N}/2=100a$) for $\gamma_{\rm N}/\Delta_0=0.2$, 1.0, and 2.0.
}
\label{fig:helical_dn}
\end{figure}

Let us now move on to the junction of the Dirac SC and DN. We consider the junction system same as that in Sec.~\ref{sec:proximity}, where  the nonmagnetic impurities are randomly distributed only in the DN region within $0< z \le L_{\rm N}$ and absent in the SC region within $-L_{\rm S}\le z \le 0$. Below, we fix $L_{\rm S}/a=400$ and $L_{\rm N}/a = 200$. The nonmagnetic impurities are incorported by the self-consistent Born approximation, where $\gamma_{\rm N}$ is the single parameter for the impurity potential (see Sec.~\ref{sec:proximity} for the detail). Figure~\ref{fig:helical_dn}(a) shows the spatial profiles of odd-frequency even-parity pair amplitudes $F^{\rm OF}(z,i\varepsilon_n)$ in the Dirac SC/DN junction, where we fix $\varepsilon_n = 0.02\Delta_0$. As expected, the odd-frequency even-parity pair amplitudes penetrating to the DN region are insensitive to the strength of nonmagnetic impurities and survives even in the diffusive limit $\gamma_{\rm N}\gg \Delta_0$. In Fig.~\ref{fig:helical_dn}(b), we plot the local DOS in the DN region ($z=L_{\rm N}/2$) for several values of $\gamma_{\rm N}$. The penetration of odd-frequency even-parity pairs is accompanied by the pronounced zero-energy peak in the local DOS at the DN region, which is referred to as the anomalous proximity effect. The peak structure of the local DOS is robust against nonmagnetic impurities. 

The robustness of the anomalous proximity effect in the Dirac SC is contrast to the proximity effect of the nematic state in STIs. As discussed in Sec.~\ref{sec:proximity}, the odd-frequency even-parity pairs in STIs penetrate into the DN region, while the amplitudes are sensitive to the strength of the nonmagnetic impurities even though the nonmagnetic impurities preserve the chiral symmetry. Hence, we attribute the fragility of the anomalous proximity effect in STIs to strong spin-orbit interaction which is essential for the topologically nontrivial property of the parent material.

\bibliographystyle{apsrev4-1_PRX_style} 
\bibliography{proximity5}

\end{document}